\PassOptionsToPackage{table,xcdraw}{xcolor}

\documentclass[acmsmall,screen]{acmart}

\AtBeginDocument{%
  }

\usepackage[most]{tcolorbox}
\usepackage{algorithm}
\usepackage{algpseudocode}
\usepackage{listings}

\usepackage{multirow}
\usepackage{tabularx}
\usepackage{threeparttable}
\usepackage{booktabs}
\usepackage{soul}
\usepackage{textcomp}

\usepackage{enumitem}
\definecolor{deepblue}{RGB}{46, 52, 64}
\definecolor{texttt}{HTML}{CC6600}
\definecolor{olivegreen}{HTML}{008800}
\definecolor{coral}{HTML}{A11F1F}
\definecolor{purple}{HTML}{4f1d8b}
\definecolor{purplegray}{HTML}{d6d6e9}
\definecolor{white1}{HTML}{4A4A4A}
\definecolor{backgroundgray}{HTML}{eef9ea}
\definecolor{backgroundgray1}{HTML}{FAF9F5}
\definecolor{framegray}{RGB}{176, 190, 197}
\definecolor{lightblue}{rgb}{0.7, 0.85, 1}
\definecolor{verylightgray}{rgb}{.97,.97,.97}

\definecolor{verylightgray}{rgb}{.97,.97,.97}

\lstdefinelanguage{Solidity}{
	keywords=[1]{anonymous, assembly, assert, balance, break, call, callcode, case, catch, class, constant, continue, constructor, contract, debugger, default, delegatecall, delete, do, else, emit, event, experimental, export, external, false, finally, for, function, gas, if, implements, import, in, indexed, instanceof, interface, internal, is, length, library, log0, log1, log2, log3, log4, memory, modifier, new, payable, pragma, private, protected, public, pure, push, require, return, returns, revert, selfdestruct, send, solidity, storage, struct, suicide, super, switch, then, this, throw, transfer, true, try, typeof, using, value, view, while, with, addmod, ecrecover, keccak256, mulmod, ripemd160, sha256, sha3}, % generic keywords including crypto operations
	keywordstyle=[1]\color{blue}\bfseries,
	keywords=[2]{address, bool, byte, bytes, bytes1, bytes2, bytes3, bytes4, bytes5, bytes6, bytes7, bytes8, bytes9, bytes10, bytes11, bytes12, bytes13, bytes14, bytes15, bytes16, bytes17, bytes18, bytes19, bytes20, bytes21, bytes22, bytes23, bytes24, bytes25, bytes26, bytes27, bytes28, bytes29, bytes30, bytes31, bytes32, enum, int, int8, int16, int24, int32, int40, int48, int56, int64, int72, int80, int88, int96, int104, int112, int120, int128, int136, int144, int152, int160, int168, int176, int184, int192, int200, int208, int216, int224, int232, int240, int248, int256, mapping, string, uint, uint8, uint16, uint24, uint32, uint40, uint48, uint56, uint64, uint72, uint80, uint88, uint96, uint104, uint112, uint120, uint128, uint136, uint144, uint152, uint160, uint168, uint176, uint184, uint192, uint200, uint208, uint216, uint224, uint232, uint240, uint248, uint256, var, void, ether, finney, szabo, wei, days, hours, minutes, seconds, weeks, years},	% types; money and time units
	keywordstyle=[2]\color{teal}\bfseries,
	keywords=[3]{block, blockhash, coinbase, difficulty, gaslimit, number, timestamp, msg, data, gas, sender, sig, value, now, tx, gasprice, origin},	% environment variables
	keywordstyle=[3]\color{violet}\bfseries,
	identifierstyle=\color{black},
	sensitive=false,
	comment=[l]{//},
	morecomment=[s]{/*}{*/},
    commentstyle=\color{black}\ttfamily,
	stringstyle=\color{red}\ttfamily,
	morestring=[b]',
	morestring=[b]"
}

\begin{document}

\title{LiquiLM: Bridging the Semantic Gap in Liquidity Flaw Audit via DCN and LLMs}

\author{Zekai Liu}
\email{zekailiu@hainanu.edu.cn}
\affiliation{%
  \institution{Hainan University}
  \city{Haikou}
  \state{Hainan}
  \country{China}
}

%% --- 第二作者 & 通讯作者 ---
\author{Xiaoqi Li}
\authornote{Corresponding author.}
\email{csxqli@ieee.org}
\affiliation{%
  \institution{Hainan University}
  \city{Haikou}
  \state{Hainan}
  \country{China}
}

%% --- 第三作者 ---
\author{Wenkai Li}
\email{cswkli@hainanu.edu.cn}
\affiliation{%
  \institution{Hainan University}
  \city{Haikou}
  \state{Hainan}
  \country{China}
}

%% --- 第四作者 ---
\author{Zongwei Li}
\email{lizw1017@hainanu.edu.cn}
\affiliation{%
  \institution{Hainan University}
  \city{Haikou}
  \state{Hainan}
  \country{China}
}

%% --- 第五作者 ---
%\author{Lei Xie}
%\email{xielei@hainanu.edu.cn}
%\affiliation{%
%  \institution{Hainan University}
%  \city{Haikou}
%  \state{Hainan}
%  \country{China}
%}

\renewcommand{\shortauthors}{Trovato et al.}

\begin{abstract}
Traditional consensus mechanisms, such as Proof of Stake (PoS), increasingly reveal an excessive dependency on large liquidity providers. Although the Proof of Liquidity (PoL) mechanism serves as a critical paradigm for incentivizing sustained liquidity provision and ensuring market stability, its transition from asset staking to active liquidity management significantly increases the complexity of underlying smart contract economic models and interaction logic. This renders hidden liquidity logic flaws difficult to detect via traditional methods, seriously threatening the system stability and user asset security of mainstream DeFi and emerging PoL ecosystems. To address this, we propose the LiquiLM framework, which integrates Large Language Models (LLMs) with a Dynamic Co-Attention Network (DCN). By establishing a dynamic interaction between liquidity-critical contracts and flaw descriptions, the framework effectively bridges the semantic gap between underlying code implementations and high-level liquidity intents. We evaluate the performance of LiquiLM on 1,490 validation contracts (covering precision, recall, specificity, and F1-score). The results show that it achieves significant effectiveness in auditing and explaining liquidity flaws: in experiments using Gemini 3 Pro and GPT-4o as backbone models, respectively, the F1-scores both exceed 90\%. Furthermore, through an in-depth audit of 1,380 real-world PoL and Ethereum economic contracts, LiquiLM successfully identifies 238 high-risk contracts and assists in discovering 10 vulnerabilities that have received CVE certification.
\end{abstract}




\maketitle

\section{Introduction}
In the DeFi ecosystem, liquidity serves not only as the cornerstone for maintaining sustained operations but also as a critical variable for ensuring market price stability \cite{li2025beyond}. While early consensus mechanisms, such as PoS, primarily prioritize the security derived from asset staking, they exhibit limitations in incentivizing sustained liquidity provision, increasingly exposing a critical dependency on large liquidity providers \cite{drossos2025automated}. To address this challenge, various Liquidity Management Paradigms emerge, with the PoL mechanism being the most representative \cite{abgaryan2024proof}. PoL aims to incentivize on-chain users to collaboratively maintain liquidity via asset staking, thereby constructing a more decentralized and robust blockchain ecosystem. However, this paradigm shift from asset staking to active liquidity management renders the underlying incentive mechanisms and interaction logic unprecedentedly intricate.

Whether for advanced PoL consensus systems or mainstream DeFi lending and trading protocols, liquidity operations heavily depend on the design and implementation of the underlying smart contracts. These contracts are tasked with orchestrating highly sensitive operations, including asset staking management, dynamic reward calculation, and value anchoring in response to market fluctuations. Consequently, any logical oversight within these contracts potentially triggers catastrophic financial losses, rendering the detection of liquidity-related flaws an imperative yet formidable task. Much like the critical need for advanced attack detection and traceability in other complex distributed systems, safeguarding the Web3 ecosystem requires highly specialized security mechanisms \cite{zhu2024sybil}.

Although academia and industry have established a relatively mature ecosystem for smart contract vulnerability detection, existing technical means remain inadequate when addressing complex liquidity flaws\cite{yang2025multi}. Current mainstream approaches primarily fall into three categories: (1) Pattern-matching-based static analysis: These methods identify known vulnerability patterns by scanning source code or bytecode \cite{grishchenko2018foundations,bu2025smartbugbert}. However, their effectiveness is heavily constrained by the completeness of the vulnerability signature database, rendering them ineffective in discovering unknown logical flaws. (2) Dynamic execution techniques: These primarily include Fuzzing and Symbolic Execution, which perform runtime detection by generating random inputs or traversing execution paths \cite{ji2023effuzz,pani2023smartfuzzdrivergen}. Nevertheless, when detecting deep state spaces, they frequently encounter dilemmas of low coverage and high computational costs caused by path explosion. (3) Formal verification: Although capable of achieving high-precision verification for specific properties via mathematical proofs \cite{almakhour2020verification}, its high algorithmic complexity and reliance on manual intervention make it difficult to scale across large-scale and variable liquidity scenarios.

More critically, unlike traditional vulnerabilities characterized by distinct syntactic patterns (such as reentrancy or arithmetic errors), Liquidity Flaws fundamentally stem from semantic deviations between low-level code implementations and high-level liquidity intents \cite{ding2025comprehensive}. Lacking fixed code signatures, these flaws are often concealed within intricate state transitions, rendering rule-based traditional tools ineffective at capturing such semantic anomalies.

As LLM technology continues to mature, numerous researchers explore its potential in the field of smart contract vulnerability auditing and explanation, achieving significant progress \cite{boi2024vulnhunt,luo2024fellmvp,sikder2025efficient}. These findings demonstrate the significant advantages of LLMs in handling complex tasks and knowledge transfer, offering new possibilities for the detection of liquidity flaws. However, in large-scale continuous contract auditing and explanation tasks, challenges such as unstable response generation, high false-positive rates, and high costs remain unresolved bottlenecks \cite{boi2024smart}. We posit that the root cause of these issues lies in two conflicting contextual dilemmas: (1) Information Scarcity: Without sufficient domain context, LLMs struggle to comprehend the logic of specific protocols, leading to off-topic responses \cite{hu2026effective, ma2025combining}; (2) Information Overload and Hallucination: When fed with massive amounts of raw code directly, excessive semantic noise interferes with the model's attention, inducing severe model hallucinations and hindering the generation of precise audit results \cite{chen2023chatgpt}.

To address these challenges, we propose LiquiLM, a framework integrating LLMs with a DCN to bridge the semantic gap in liquidity flaw detection. This framework achieves precise localization and explanation of liquidity flaws through the collaboration of three core modules: First, to eliminate semantic noise within the code, we perform standardization and slicing on the target contract to generate refined code embedding vectors. Second, to provide LLMs with high-quality, low-noise prior knowledge, we introduce the DCN module to establish a deep semantic interaction alignment between the contract slice vectors and the liquidity feature description vectors. This process accurately identifies and filters out the vast majority of safe slices irrelevant to liquidity, thereby significantly reducing the auditing load on LLMs. Finally, to mitigate the uncontrollability of LLMs generation, we design a Four-Phase Collaborative Prompt System to guide LLMs in performing targeted auditing within a strictly constrained problem space. Compared to existing tools, LiquiLM captures liquidity-related operational details and subtle code nuances more acutely, significantly enhancing the capability to detect hidden flaws across diverse scenarios, while providing interpretable, customized audit reports and remediation suggestions.

The main contributions of this paper are as follows:
\begin{itemize}
    \item To our best knowledge, we are the first to propose the LiquiLM framework, which utilizes LLMs combined with DCN technology to analyze and optimize liquidity flaws in PoL smart contracts.

    \item We design a Four-Phase Collaborative Prompt System that guides LiquiLM through step-by-step analysis of complex smart contracts, verifying accuracy at each stage and significantly enhancing the quality and reliability of the final audit report.

    \item Empirical Validation and High Reliability. Through an extensive evaluation of 1,490 real-world contracts, LiquiLM demonstrates superior detection efficacy. Powered by Gemini 3 Pro, it achieves a recall exceeding 86\% and an F1-score above 84\% across both liquidity flaws and traditional vulnerabilities.
%    \item The code and experimental data for LiquiLM have been made publicly available at the following anonymous link: \url{https://figshare.com/s/3bfe12f83fa35ed301c3}.
\end{itemize}
\section{Background}

\subsubsection*{\textbf{Large Language Models}}
LLMs have evolved into advanced deep learning frameworks capable of capturing complex semantic relationships within vast corpora. Representative models, such as the GPT series and Google's Gemini, have demonstrated exceptional performance not only in natural language tasks like summarization \cite{ding2024evaluation} and translation \cite{huang2023towards} but also in code-related domains. By training on massive datasets containing multi-language source code, these models can comprehend the syntax and logic of programming languages like Solidity and Python, assisting developers in code generation, optimization, and vulnerability detection \cite{nam2024using, huang2023bias}.

Notably, models like Gemini utilize advanced architectures to support long-context understanding and complex reasoning, enabling zero-shot and few-shot learning capabilities \cite{deng2024wav2prompt}. However, despite these advancements, applying LLMs directly to high-stakes auditing tasks remains challenging. Without sufficient domain constraints, they are prone to generating erroneous or misleading outputs (hallucinations), particularly when interpreting highly specialized logic \cite{perkovic2024hallucinations}.

\subsubsection*{\textbf{The Proof-of-Liquidity Mechanism}}
Different from traditional mechanisms such as PoS, which prioritize asset staking for security, the PoL mechanism algorithmically integrates consensus security with liquidity provision \cite{abgaryan2024proof}. Under this paradigm, token incentives are dynamically bound to the depth and duration of user-provided liquidity. This shift introduces a complex economic interaction layer where users are incentivized not merely to hold assets but to actively participate in market making and liquidity pool operations.

Currently, several emerging platforms, including Berachain, experiment with the PoL mechanism to optimize liquidity management. However, while the PoL design successfully fosters significant diversification and decentralization of liquidity providers, it also notably increases the complexity of the underlying system. The interaction logic shifts from simple state storage to continuous, dynamic value calculation. The complex coupling between consensus rewards and market behaviors forms a vast state space, where subtle logical inconsistencies in incentive algorithms potentially trigger liquidity manipulation or drainage, rendering traditional vulnerability detection methods ineffective.

\subsubsection*{\textbf{Liquidity-Critical Smart Contracts}}
Liquidity-Critical Smart Contracts serve as the executable infrastructure for PoL systems and DeFi protocols \cite{john2023smart}. Unlike general-purpose logic contracts, these contracts are tasked with orchestrating sensitive financial primitives that directly impact market stability. Key operational characteristics include:
\begin{itemize}
    \item \textbf{Staking Lifecycle Control:} Managing the precise state transitions of user assets, including locking, unlocking, and slashing conditions.
    \item \textbf{Dynamic Reward Distribution:} Real-time calculation of incentives based on fluctuating liquidity contributions and protocol-defined weights.
    \item \textbf{Liquidity Flow Management:} Automating the routing and transfer of liquidity between different pools or users without manual intervention.
    \item \textbf{Value Peg Maintenance:} Executing algorithmic adjustments to maintain asset price stability in response to market volatility.
\end{itemize}

\subsubsection*{\textbf{Liquidity Flaws}}

In the design and implementation of liquidity-critical smart contracts, specific design flaws potentially induce severe liquidity security issues, including abnormal outflow, asset value de-pegging, and liquidity depletion. We collectively term these fundamental flaws, which threaten the stability and fairness of contract liquidity, as Liquidity Flaws.

Unlike traditional smart contract vulnerabilities, Liquidity Flaws fundamentally stem from the failure of contract logic to enforce security invariants under specific conditions. Based on the source of state dependency, we classify these flaws into two categories \cite{wu2025security,wu2025exploring}:

\begin{itemize}
    \item \textbf{Endogenous Flaws}: These originate from the contract's internal logic implementation and governance mechanisms, specifically including Logic Implementation Flaws, Business Protocol Flaws, and Governance Authority Risks.
    
    \item \textbf{Exogenous Flaws}: These originate from unsafe dependencies on external data inputs or trust assumptions, specifically including Liquidity Valuation Distortion and Transient Liquidity Shock.
\end{itemize}

Table \ref{tab:liquidity_flaws} details our formal definitions and classification of the various Liquidity Flaws.

\definecolor{grayhead}{gray}{0.95}

\begin{table}[htbp]
\centering
\caption{Taxonomy of Smart Contract Liquidity Flaws}
\label{tab:liquidity_flaws}
\resizebox{\textwidth}{!}{ % 开始缩放
\renewcommand{\arraystretch}{1} % 调整行间距
\footnotesize 
% 假设你已经定义了 grayhead 颜色，如果没有，请取消下面这行的注释
% \definecolor{grayhead}{gray}{0.9}

\begin{tabular}{@{} l l >{\raggedright\arraybackslash}p{9.1cm} @{}} 
\toprule
% 表头：使用 multicolumn{1}{c}{...} 强制居中
\rowcolor{grayhead} 
\multicolumn{1}{c}{\textbf{Abbr.}} & 
\multicolumn{1}{c}{\textbf{Category}} & 
\multicolumn{1}{c}{\textbf{Definition / Mechanism}} \\ 
\midrule

% --- Group I ---
\rowcolor{grayhead} 
\multicolumn{3}{c}{\textit{\textbf{Group I: Endogenous Flaws}}} \\ \addlinespace[3pt]
\textbf{LIF} & Logic Implementation Flaws & Execution-level errors causing state-model divergence. \\
\textbf{BPF} & Business Protocol Flaws & Economic design flaws causing unintended leakage without syntax violation. \\
\textbf{GAR} & Governance Authority Risks & Systemic risks from compromised control, enabling unauthorized asset drainage. \\ 
\midrule

% --- Group II ---
\rowcolor{grayhead} 
\multicolumn{3}{c}{\textit{\textbf{Group II: Exogenous Flaws}}} \\ \addlinespace[3pt]
\textbf{LVD} & Liquidity Valuation Distortion & On-chain value/market price misalignment due to manipulated or delayed data. \\
\textbf{TLS} & Transient Liquidity Shock & Market equilibrium disruption via massive atomic capital. \\ 
\bottomrule
\end{tabular}
}
\end{table}

\subsubsection*{\textbf{Dynamic Co-attention Network}}
The DCN is a deep evolutionary architecture of the Co-attention Mechanism \cite{wang2025knowledge}. The Co-attention Mechanism has widely demonstrated its superiority in the field of Natural Language Processing (NLP) due to its ability to effectively capture fine-grained interaction features between queries and documents \cite{nelatoori2025toxic}. Building upon this foundation, DCN introduces a multi-layer iterative update mechanism, which significantly enhances the model's capacity for deep semantic understanding of complex input information.

Within the LiquiLM, we incorporate DCN as a core pre-filtering module. By performing bidirectional semantic alignment between the target contract slices and liquidity flaw descriptions, DCN accurately identifies and filters out over 90\% of safe slices. This mechanism significantly compresses the search space for the LLM, effectively mitigating the high cost and latency bottlenecks common in large-scale contract auditing tasks. Furthermore, the semantic flow information captured by DCN provides rich contextual cues for the LLMs. Serving as semantic anchors, these cues transform traditional comprehensive scanning into targeted, precise auditing, thereby further reducing computational overhead while suppressing model hallucinations \cite{gunjal2024detecting}.

\begin{figure*}[h!]
\centering
\includegraphics[width=\textwidth]{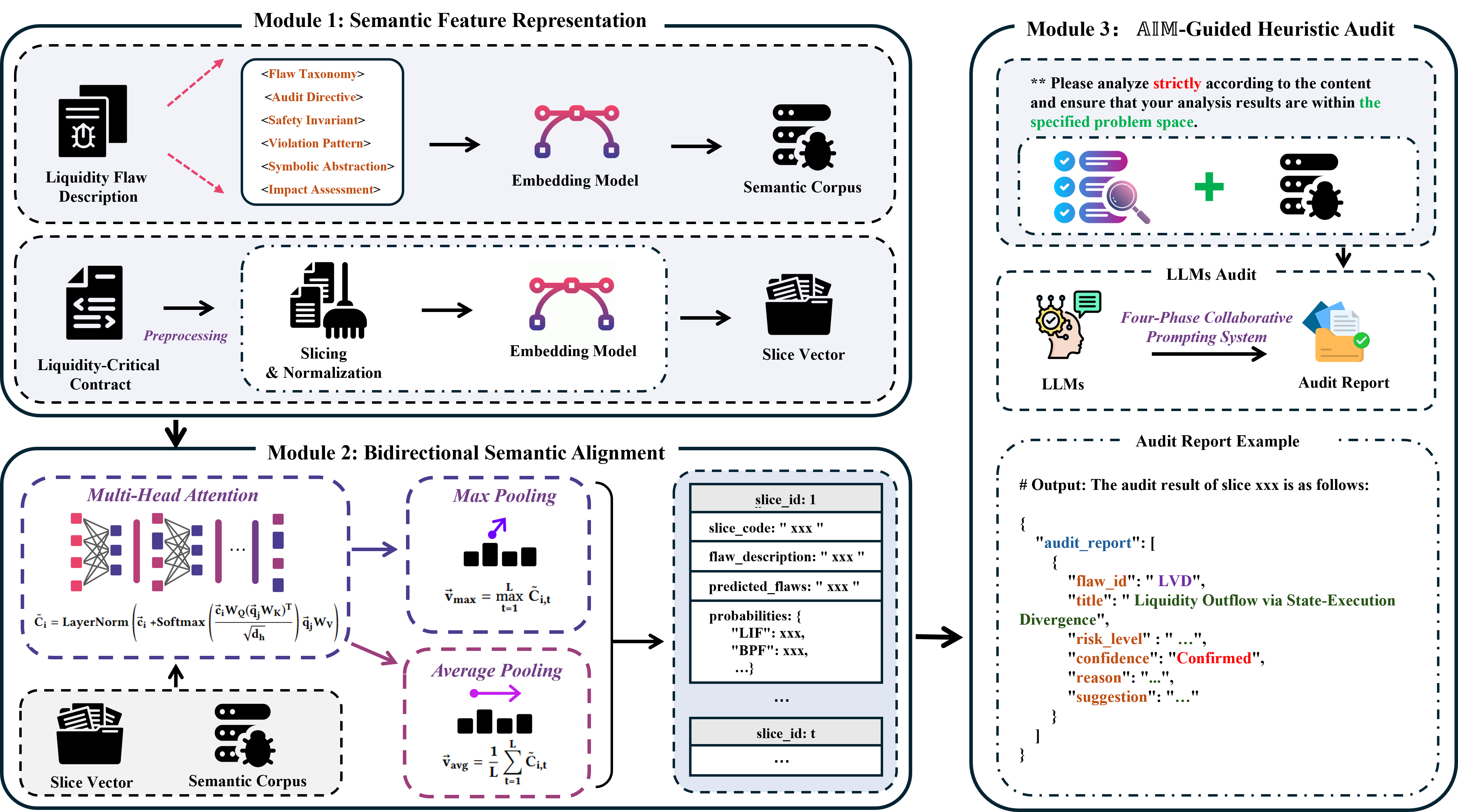}
\caption{\textbf{The Overall Architecture of LiquiLM.} \textit{Note}: The \textbf{Semantic Feature Representation} module slices and normalizes the target liquidity-critical contract source code to generate embedding vectors, while simultaneously constructing a liquidity defect semantic corpus. The \textbf{Bidirectional Semantic Alignment} module employs a DCN model to align contract slice vectors with corpus entries; following max pooling and average pooling, it generates the Audit-Informed Manifest ($\mathbb{AIM}$). Finally, the \textbf{$\mathbb{AIM}$-Guided Heuristic Audit} module adopts a Four-Phase Collaborative Prompt System to guide LLMs in performing an in-depth analysis of critical slices within the $\mathbb{AIM}$ and generating the final audit report.}
\label{fig:overall}
\end{figure*}
\section{LiquiLM}

The LiquiLM framework, as shown in Fig.~\ref{fig:overall}, consists of three modules: In \underline{\textbf{Module 1}}, the framework transforms raw, unstructured Solidity source code into semantically deep embedding vectors understandable by the DCN model. It also maintains a specialized PoL Semantic Corpus as an auditing benchmark. In \underline{\textbf{Module 2}}, the framework uses DCN technology to achieve interactive alignment between features and knowledge. This process generates an $\mathbb{AIM}$, which encapsulates code logic and flaw details mapped within the semantic space. In \underline{\textbf{Module 3}}, $\mathbb{AIM}$ serves as a guide for LLMs, enhancing contextual understanding through a Four-Phase Collaborative Prompt System that executes planning, verification, and feedback calibration. This enables the system to generate high-quality structured audit reports that balance both precision and recall.

\subsection{Module 1: Semantic Feature Representation}

Module 1 processes two independent information sources: a collection of target PoL contract source codes $I_1$ and a set of liquidity flaw descriptions $I_2$. Let $I_1 = \{C_1, C_2, \ldots, C_n\}$ denote the set of source codes. Let $I_2 = \{Q_1, Q_2, \ldots, Q_m\}$ represent the set of liquidity flaw type descriptions, where each $Q_i = \{q_1, q_2, \ldots, q_k\}$ corresponds to the specific flaw category.

\subsubsection*{\textbf{Dependency-aware Slicing and Normalization}}

Each contract source code $C_i$ is defined as a union of a global variable set $\mathcal{G}$, a modifier set $\mathcal{M}$, and a function set $\mathcal{F}$. After removing noise such as comments and empty lines through preprocessing, a slicing operation is performed for every $f_{i} \in \mathcal{F}$ as defined in Eq.\eqref{eq:1}:

\begin{equation}
\label{eq:1}
\begin{cases}
\mathcal{M}{dep} = { m \in M \mid \text{name}(m) \in \text{Tokens}(f{i}) } 

\\mathcal{G}{dep} = { g \in \mathcal{G} \mid \text{Vars}(g) \cap \text{Tokens}(f{i} \cup \mathcal{M}{dep}) \neq \emptyset } 

\\S(f{i}) = { f_{i} } \cup \mathcal{G}{dep} \cup \mathcal{M}{dep}
\end{cases}
\end{equation}

where $\mathcal{G}_{dep}$ and $\mathcal{M}_{dep}$ represent the dependency sets of $f_{i}$. 

This slicing process retains only the context directly related to the functional logic flow, thereby minimizing noise interference.

To eliminate naming bias in variable and function names that may adversely affect the DCN model, a normalization function is applied to transform identifier sequences into abstract placeholders. This ensures the model focuses on structural logic rather than lexical vocabulary. Simultaneously, protected structural tags are defined to preserve annotation fields such as \texttt{[Target\_Function]}, ensuring the model correctly identifies the hierarchical levels of the slice.

\subsubsection*{\textbf{Embedding Vector Generation}}
To convert the information sources $I_1$ and $I_2$ into vector representations, we employ the pre-trained encoder BGE-M3 \cite{chen2025m3embeddingmultilingualitymultifunctionalitymultigranularity}. This process is expressed as Eq.\eqref{eq:2}:

\begin{equation}
\label{eq:2}
\mathbf{H} = E(I) \in \mathbb{R}^{L \times D}
\end{equation}

where $L$ is the sequence length (512) and $D$ is the feature dimension (1024). For $I_1$ and $I_2$, we obtain the following vector sets:

\begin{itemize}
\item $\vec{I}_1 = \{\vec{C}_i \mid \vec{C}_i = \{\vec{c}_1, \vec{c}_2, \ldots, \vec{c}_t\}, \forall i \in \{1, 2, \ldots, n\}\}$
\item $\vec{I}_2 = \{\vec{Q}_i \mid \vec{Q}_i = \{\vec{q}_1, \vec{q}_2, \ldots, \vec{q}_k\}, \forall i \in \{1, 2, \ldots, m\}\}$
\end{itemize}The resulting set $\vec{I}_2$ constitutes the \textbf{Semantic Corpus}.

\subsection{Module 2: Bidirectional Semantic Alignment}

The DCN module is designed to construct a deep semantic mapping between code slices and liquidity flaw descriptions, yielding the Audit-Informed Manifest $\mathbb{AIM}$ as its output. Serving as a prior input for downstream LLMs auditing, $\mathbb{AIM}$ efficiently filters out the vast majority of irrelevant safe slices in $\vec{I}_1$, guiding the audit process to precisely focus on specific flaw patterns described in $\vec{I}_2$.

Distinct from shallow metric methods such as cosine similarity, DCN achieves fine-grained semantic alignment through a multi-layer co-attention mechanism. For any given code slice vector $\vec{c}_i \in \mathbb{R}^{L \times D}$ and target flaw description vector $\vec{q}_j \in \mathbb{R}^{M \times D}$, the complete interaction and inference process is described by Algorithm \ref{alg:dcn_inference}, covering three key stages: semantic projection and interaction, dual-channel feature aggregation and prediction, and a cost-sensitive optimization objective.

% Algorithm 1: DCN Forward Propagation
\begin{algorithm}
\caption{DCN Forward Propagation for Semantic Alignment}
\label{alg:dcn_inference}
\begin{algorithmic}[1]
\Require 
  Code Slice Embedding $\vec{C} \in \mathbb{R}^{L \times D}$ (from $\vec{I}_1$), 
  Target Flaw Description $\vec{Q} \in \mathbb{R}^{L_q \times D}$ (from $\vec{I}_2$)
\Ensure Independent Confidence Score $p \in [0, 1]$

\State \textbf{Step 1: Semantic Projection}
\State $H_C \leftarrow \text{LayerNorm}(W_c \vec{C} + b_c)$ \Comment{Map to latent dim $d_h=256$}
\State $H_Q \leftarrow \text{LayerNorm}(W_q \vec{Q} + b_q)$

\State \textbf{Step 2: Query-Guided Co-Attention}
\For{$l \gets 1$ to $N$}
    \State \Comment{Code serves as Query, Flaw Description as Key/Value}
    \State $H_{attn} \leftarrow \text{MultiHeadAttn}(Q=H_C, K=H_Q, V=H_Q)$
    \State $H_C \leftarrow \text{LayerNorm}(H_C + \text{Dropout}(H_{attn}))$ 
    \State $H_{ffn} \leftarrow \text{FFN}(H_C)$
    \State $H_C \leftarrow \text{LayerNorm}(H_C + \text{Dropout}(H_{ffn}))$
\EndFor

\State \textbf{Step 3: Dual-Channel Aggregation}
\State $v_{max} \leftarrow \max_{t=1}^{L} H_{C,t}$ \Comment{Global Max Pooling (Eq. 4)}
\State $v_{avg} \leftarrow \frac{1}{L} \sum_{t=1}^{L} H_{C,t}$ \Comment{Global Avg Pooling (Eq. 5)}
\State $\vec{v}_{final} \leftarrow \text{Concat}(v_{max}, v_{avg})$

\State \textbf{Step 4: Classification}
\State $p \leftarrow \sigma(\text{MLP}(\vec{v}_{final}))$ \Comment{Sigmoid activation (Eq. 6)}
\State \Return $p$
\end{algorithmic}
\end{algorithm}

\subsubsection*{\textbf{Semantic Projection and Interaction}}
To suppress redundant noise in the high-dimensional feature space and align heterogeneous semantics, we first map the input $\vec{c}_i$ and flaw description vector $\vec{q}_j$ from the original dimension $D=1024$ to a latent feature space $d_h=256$ via learnable linear projection layers.

Subsequently, the model adopts a Query-Guided Co-Attention Mechanism to dynamically enhance slice features. Specifically, the model inputs $\vec{c}_i$ as the Query, and the flaw description vector $\vec{q}_j$ as the Key and Value into a multi-head attention network. Under this mechanism, the model computes the semantic correlation between each token in the slice and the flaw description, and utilizes residual connections to generate the enhanced representation $\tilde{\mathbf{C}}_i$, as shown in Eq.\eqref{eq:3}:

\begin{equation}
\label{eq:3}
\tilde{\mathbf{C}}_i = \text{LayerNorm}\left(\vec{c}_i + \text{Softmax}\left(\frac{\vec{c}_i W_Q (\vec{q}_j W_K)^T}{\sqrt{d_h}}\right) \vec{q}_j W_V\right)
\end{equation}

where $W_Q, W_K, W_V$ represent learnable projection matrices. If the local logic flow in the slice exhibits high semantic correlation with the flaw features described by $\vec{q}_j$, the attention weights of the corresponding regions are significantly amplified. Through multi-layer interaction, $\vec{c}_i$ is reconstructed into an enhanced representation $\tilde{\mathbf{C}}_i$ fused with specific flaw contexts.

\subsubsection*{\textbf{Dual-Channel Aggregation and Prediction}}
Although the interactive slice representation $\tilde{\mathbf{C}}_i$ incorporates flaw features, it remains a variable-length token sequence (retaining the sequence dimension $L$). To aggregate these fine-grained local semantic features into a fixed-dimensional global representation while capturing significant flaw-triggering features and global semantic context, we apply a dual-channel pooling strategy along the sequence dimension:

\begin{itemize}
    \item \textbf{Global Max Pooling (GMP)}: Designed to capture local saliency. In the slice sequence of length $L$, GMP filters out the feature units with the highest response values to the flaw description vector $\vec{q}_j$, thereby locking onto potential flaw trigger points while ignoring irrelevant code interference, as expressed in Eq.\eqref{eq:4}:
    \begin{equation}
    \label{eq:4}
    \vec{v}_{\max} = \max_{t=1}^{L} \, \tilde{\mathbf{C}}_{i,t}
    \end{equation}
    
    \item \textbf{Global Avg Pooling (GAP)}: Designed to extract global semantic logic. GAP smooths the slice sequence to reflect the overall functional intent of the code slice, providing necessary background context for discrimination, as expressed in Eq.\eqref{eq:5}:
    \begin{equation}
    \label{eq:5}
    \vec{v}_{\text{avg}} = \frac{1}{L} \sum_{t=1}^{L} \tilde{\mathbf{C}}_{i,t}
    \end{equation}
\end{itemize}

%Finally, by concatenating the outputs of GMP and GAP, we construct a joint feature vector that considers both local flaw features and global logic. This vector is then input into a Multi-Layer Perceptron (MLP) to generate the prediction probability vector $P_i = \{p_1, p_2, \dots, p_K\}$, as defined in Eq.\eqref{eq:6}:

Finally, by concatenating the outputs of GMP and GAP, we construct a comprehensive joint feature vector that effectively captures both local flaw features and global logic. This unified vector is then fed into a Multi-Layer Perceptron (MLP) to produce the prediction probability vector $P_i = \{p_1, p_2, \dots, p_K\}$, as defined in Eq.\eqref{eq:6}:

\begin{equation}
\label{eq:6}
P_i = \sigma\left(\text{MLP}(\vec{v}_{\max} \oplus \vec{v}_{\text{avg}})\right)
\end{equation}

where $\oplus$ denotes the vector concatenation operation, and $\sigma$ is the Sigmoid activation function. In this definition, $p_k \in [0, 1]$ represents the independent confidence for the presence of the $k$-th flaw.

% Algorithm 2: Training Strategy
\begin{algorithm}
\caption{Cost-Sensitive Training with Semantic Corpus}
\label{alg:training_strategy}
\begin{algorithmic}[1]
\Require 
  Training Set $\mathcal{D} = \{(\vec{C}_i, Y_i)\}_{i=1}^{M}$ where $Y_i \in \{0,1\}^K$,
  Semantic Corpus $\vec{I}_2 = \{\vec{Q}_1, \dots, \vec{Q}_K\}$,
  Imbalance Weight $\alpha=6.0$, 
  Learning Rate $\eta$
\Ensure Optimized Model Parameters $\theta^*$

\State Initialize parameters $\theta$ randomly
\For{$epoch \gets 1$ to $E_{max}$}
    \For{each batch $\mathcal{B} \subset \mathcal{D}$}
        \State $\mathcal{L}_{batch} \leftarrow 0$
        
        \For{each sample $(\vec{C}_i, Y_i)$ in $\mathcal{B}$}
            \State $\mathcal{L}_{sample} \leftarrow 0$
            
            \Comment{Iterate over all $K$ flaw types (Multi-label Strategy)}
            \For{$k \gets 1$ to $K$}
                \State \textbf{Forward Pass:}
                \State $p_{i,k} \leftarrow \text{DCN\_Forward}(\vec{C}_i, \vec{Q}_k; \theta)$ 
                
                \State \textbf{Compute Weighted Loss (Eq. 7):}
                \If{$Y_{i,k} = 1$} \Comment{Ground Truth: Flaw exists}
                    \State $\ell_k \leftarrow - \alpha \cdot \log(p_{i,k})$ 
                \Else \Comment{Ground Truth: Safe / Noise}
                    \State $\ell_k \leftarrow - \log(1 - p_{i,k})$
                \EndIf
                \State $\mathcal{L}_{sample} \leftarrow \mathcal{L}_{sample} + \ell_k$
            \EndFor
            
            \State $\mathcal{L}_{batch} \leftarrow \mathcal{L}_{batch} + (\mathcal{L}_{sample} / K)$
        \EndFor
        
        \State $\mathcal{L}_{batch} \leftarrow \mathcal{L}_{batch} / |\mathcal{B}|$
        \State Update parameters: $\theta \leftarrow \theta - \eta \cdot \nabla_\theta \mathcal{L}_{batch}$
    \EndFor
\EndFor
\State \Return $\theta^*$
\end{algorithmic}
\end{algorithm}

\subsubsection*{\textbf{Cost-Sensitive Optimization Objective}}
Given that the slice samples containing flaws in $\vec{I}_1$ are extremely sparse (i.e., a severe imbalance between positive and negative samples), standard loss functions tend to induce the model to overfit to the majority class (safe slices). To address this, we introduce the Weighted Binary Cross-Entropy Loss \cite{xie2015holistically}, as shown in Eq.\eqref{eq:7}:

\begin{equation}
\label{eq:7}
\mathcal{L} = - \frac{1}{K} \sum_{k=1}^{K} \left[ \alpha \cdot y_k \log p_k + (1 - y_k) \log (1 - p_k) \right]
\end{equation}

By introducing the penalty coefficient $\alpha$ (set to 6.0 in experiments), we significantly increase the penalty weight for False Negatives, thereby effectively mitigating the gradient dominance effect caused by the disparity between positive and negative samples. This strategy forces the DCN to prioritize sparse flaw patterns during optimization, maximizing the recall rate of flaw slices while guaranteeing precision. Algorithm \ref{alg:training_strategy} shows the complete cost-sensitive training process based on a semantic corpus.

Compared to directly inputting the raw information source $\vec{I}_1$ into LLMs, utilizing the DCN to learn semantic flows yields an $\mathbb{AIM}$ that provides richer and higher-quality context for LLMs, thereby enhancing their performance in large-scale liquidity flaw auditing tasks. Specifically, the Semantically Aligned Slices enable LLMs to capture and integrate existing flaw features more efficiently, while the Flaw Prediction Probability Vectors $P_i$ serve as credibility references, assisting LLMs in verifying and prioritizing flaws.

\subsection{Module 3: $\mathbb{AIM}$-Guided Heuristic Audit}
\subsubsection*{\textbf{Two-Stage Confidence Filtering Strategy}}
To optimize the information quality of inputs to LLMs, we introduce a Two-Stage Confidence Filtering Strategy during the construction of $\mathbb{AIM}$. This strategy aims to filter out background noise while effectively identifying and retaining potential flaws that exhibit low confidence but salient features, thereby reducing false negatives.

First, based on the independent probability vector $P_i = \{p_1, ..., p_K\}$ output by the DCN model, we employ the F1-optimal threshold $\tau_{high}=0.3$ derived from the validation set and a noise threshold $\tau_{low}=0.001$ to perform an initial partition of the prediction results:

\begin{itemize}
    \item \textbf{Absolute Confidence Interval ($p_k \ge \tau_{high}$)}: Classified as high risk and directly retained;
    \item \textbf{Noise Discard Interval ($p_k < \tau_{low}$)}: Regarded as background noise and eliminated;
    \item \textbf{Fuzzy Confidence Interval ($\tau_{low} \le p_k < \tau_{high}$)}: Exhibits epistemic uncertainty and requires secondary evaluation.
\end{itemize}

Targeting the fuzzy confidence interval, we propose a Peak-to-Noise Ratio (PNR) mechanism to enhance discriminative power. PNR is defined as the ratio of the target probability $p_k$ to the background noise mean $\mu_{\text{noise}}$, calculated as shown in Eq.~\eqref{eq:8} to~\eqref{eq:9}:

\begin{equation}
\label{eq:8}
\mu_{\text{noise}} = \frac{1}{|N|} \sum_{j \in N} p_j, \quad \text{where } N = \{j \mid p_j < \tau_{low}\}
\end{equation}

\begin{equation}
\label{eq:9}
\text{PNR}(k) = \frac{p_k}{\mu_{\text{noise}} + \epsilon}
\end{equation}

We set $\tau_{pnr}=10$, requiring $p_k$ to be at least one order of magnitude higher than the background noise to be retained. This strategy fully leverages the independence characteristic of the Sigmoid output and, compared to traditional normalization methods, effectively mitigates the mutual suppression risk potentially triggered in multi-flaw co-occurrence scenarios.

\subsubsection*{\textbf{Four-phase Collaborative Prompting System}}

\begin{figure}[h!]
  \centering
  \includegraphics[width=\linewidth]{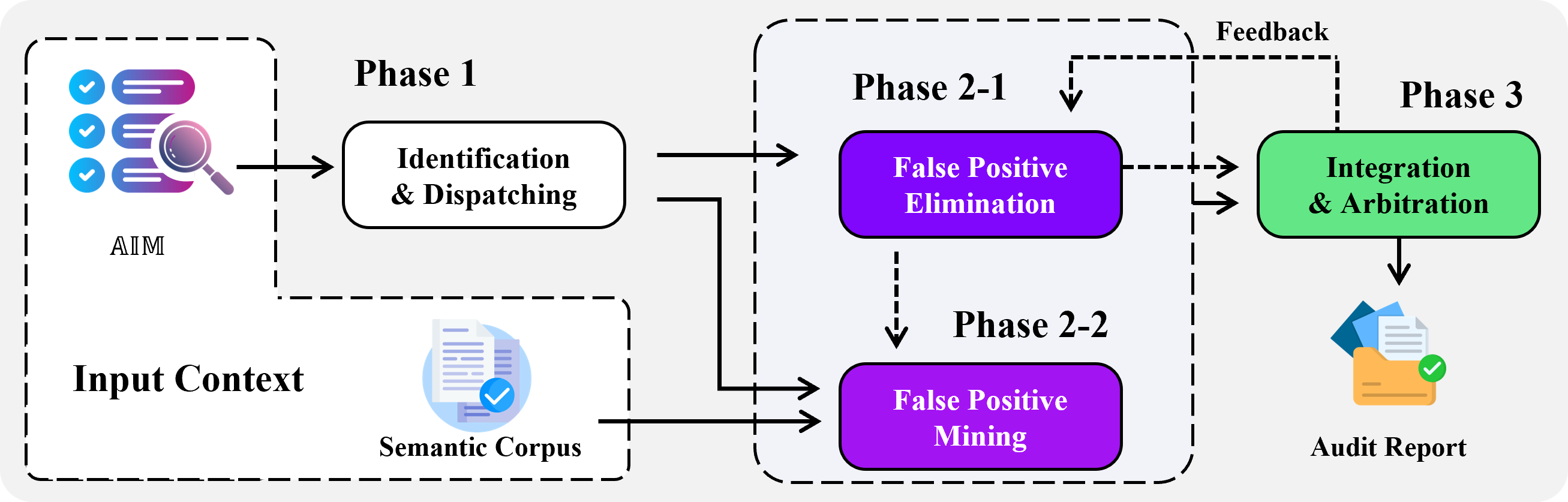}
  \caption{Four-Phase Collaborative Prompt System of LiquiLM.}
  \label{fig: prompt}
\end{figure}

We integrate the updated $\mathbb{AIM}$ and Semantic Corpus $I_2$ as the input context for LLMs. To ensure the stability and standardization of model responses, we design a Four-Phase Collaborative Prompt System. Fig.~\ref{fig: prompt} illustrates the overall architecture of this system, clearly presenting the workflow and data flow relationships of each phase.

\paragraph{\textbf{\texttt{Phase 1: Identification \& Dispatching}}}

Phase 1 serves as the entry point of the system, responsible for the lightweight parsing and task dispatching of $\mathbb{AIM}$. Without introducing deep analysis, this phase extracts probability vector information from $\mathbb{AIM}$ and performs routing: 
\begin{itemize}
\item if the current audit slice $c_i$ contains a probability vector $p_k$ belonging to the absolute confidence interval, it extracts the relevant context and routes it to Phase 2-1 for verification; 
\item if it only contains vectors belonging to the fuzzy confidence interval, the slice is routed to Phase 2-2 for deep mining. This phase ultimately outputs standardized parsing results in JSON format.
\end{itemize}

\paragraph{\textbf{\texttt{Phase 2-1: False Positive Elimination}}}

This phase focuses on reviewing potential flaws proposed by $\mathbb{AIM}$ (forwarded via Phase 1) or Phase 2-2, aiming to eliminate false positives through precise analysis. It comprises two processing modes and generates a JSON report containing judgment rationales:

\begin{itemize}
\item \textbf{Initial Review}: When the input originates from Phase 1, this phase conducts a refined screening of high-confidence flaws flagged by $\mathbb{AIM}$. If Phase 2-1 determines that the flaw is valid, it records detailed audit information and passes it to Phase 3; if determined as a false positive (i.e., an $\mathbb{AIM}$ prediction error), the slice undergoes downgrade processing and is transferred to Phase 2-2 for secondary exploration.
\item \textbf{Feedback Verification}: When the input originates from the feedback loop of Phase 3, this phase conducts cross-verification based on audit details provided by Phase 2-2 to confirm the authenticity of its findings, and returns the final verification results to Phase 3.
\end{itemize}

\paragraph{\textbf{\texttt{Phase 2-2: False Negative Mining}}}

Phase 2-2 aims to discover hidden flaws missed by $\mathbb{AIM}$ or excluded by Phase 2-1. This phase receives fuzzy samples from Phase 1 and excluded samples from Phase 2-1, utilizing the complete set of liquidity flaw descriptions $I_2$ as guidance to conduct a comprehensive breadth-wise scan. It records all newly discovered potential flaws and their audit details as structured JSON and submits them to Phase 3 for aggregation.

\paragraph{\textbf{\texttt{Phase 3: Integration \& Arbitration}}}

This phase is responsible for the logical integration and conflict arbitration of outputs from Phase 2-1 and Phase 2-2. The system introduces an iterative feedback mechanism: if Phase 2-2 detects a new flaw not recorded by Phase 2-1, or if a divergence exists in the judgment of the same flaw between the two phases, Phase 3 triggers the feedback loop, returning the disputed item to Phase 2-1 for re-evaluation. Finally, the system generates a unified JSON audit report. 

The report adopts a strict labeling strategy: flaws are marked as \texttt{[Confirmed]} only when they pass the Phase 2-1 review or are finally verified through the feedback verification loop; flaws identified solely by Phase 2-2 but not passed through cross-verification are marked as \texttt{[Suspicious]}.

This four-stage audit process enables LLMs to conduct progressive analysis, advancing gradually from simple to complex and from macro to micro levels. Compared to traditional single-prompt methods, this architecture significantly reduces the error rate and enhances the ability to focus on specific flaws. In particular, the feedback loop between Phase 3 and Phase 2-1 is crucial; it establishes an adversarial detection mechanism between Phase 2-1 and Phase 2-2, effectively mitigating missing detections caused by insufficient coverage of $\mathbb{AIM}$, while also suppressing false detection problems induced by information overload or model hallucinations in Phase 2-2.

\section{Assessment}
In the evaluation section, we focus on the following four Research Questions (RQs) to comprehensively validate the effectiveness and robustness of our proposed framework: \underline{RQ1:} Can the DCN effectively mitigate the propagation of errors to downstream tasks at the source? \underline{RQ2:} How do the $\mathbb{AIM}$ and the Four-Phase Collaborative Prompt System contribute to the performance enhancement of LiquiLM? \underline{RQ3:} Does LiquiLM demonstrate superior performance compared to existing state-of-the-art tools? \underline{RQ4:} Is LiquiLM effective in auditing liquidity flaws within uncontrolled real-world scenarios?

To answer these questions, we designed four targeted experiments: \underline{Exp.1 (for RQ1):} Evaluates the performance of DCN during both training and inference phases to verify its reliability as a preliminary filter. \underline{Exp.2 (for RQ2):} Assesses the effectiveness of LiquiLM in auditing liquidity flaws, including an ablation study on the construction of $\mathbb{AIM}$ and the Four-Phase Collaborative Prompt System. \underline{Exp.3 (for RQ3):} Conducts a comparative analysis between LiquiLM and baseline techniques as well as existing automated tools. \underline{Exp.4 (for RQ4):} Applies LiquiLM to audit liquidity-critical smart contracts in real-world scenarios to validate its practical utility.

\subsection*{\textbf{Dataset Construction}}

\subsubsection*{\textbf{Data Collection}}

Our dataset primarily consists of the following four components:

\begin{itemize}
\item $I_{1(\text{Core})}$: Targeting Exp.1 and Exp.2, we screen and collect 650 smart contracts containing liquidity flaws from nine public datasets, including the OWASP Smart Contract Top 10, generating a total of 7,144 contract slices.
\item $I_{1(\text{Baseline})}$: Targeting Exp.3, we collect 840 contracts with traditional vulnerabilities from the SCV-1-2000 dataset, generating a total of 9,011 contract slices.
\item $I_{1(\text{Real})}$: Targeting Exp.4, we collect 1,380 liquidity-critical smart contracts from Berachain (based on the PoL) and the Ethereum Mainnet, generating a total of 12,538 contract slices.
\item $I_2$: Comprises a repository of flaw/vulnerability descriptions compiled by a combination of manual filtering and AI-assisted generation.
\end{itemize}

%Notably, when constructing $I_{1(\text{core})}$ and $I_{1(\text{baseline})}$, we do not introduce completely flaw-free benign contracts. Instead, we employ Contract Slicing techniques to construct positive and negative sample sets. This method effectively guarantees fine-grained alignment of samples. The specific composition of $I_{1(\text{core})}$ and $I_{1(\text{baseline})}$ is detailed in Table \ref{tab:dataset_side_by_side}.

Notably, we constructed $I_{1(\text{core})}$ and $I_{1(\text{baseline})}$ using contract slicing to ensure fine-grained alignment between positive and negative samples. To rigorously prevent data leakage, we strictly enforced a contract-level splitting strategy for both training and evaluation. Table~\ref{tab:dataset_side_by_side} details the specific composition of these datasets.

\begin{table}[htbp]
\centering
\caption{\textbf{Detailed Composition of Datasets $I_{1(\text{core})}$ and $I_{1(\text{baseline})}$.}}
\label{tab:dataset_side_by_side}
\renewcommand{\arraystretch}{1.2}
\scriptsize % 稍微改小字体以适应并排
\setlength{\tabcolsep}{3pt} % 稍微减小列间距

\begin{tabular}{@{\extracolsep{\fill}} llrr  llrr @{}} % 移除了中间的 |
\toprule
%\rowcolor{grayhead} 
% 移除了 \multicolumn 中的 |
\multicolumn{4}{c}{\textbf{$I_{1(\text{Core})}$: Liquidity Flaws}} & 
\multicolumn{4}{c}{\textbf{$I_{1(\text{Baseline})}$: Traditional Vulnerabilities}} \\ 
\cmidrule(r){1-4} \cmidrule(l){5-8} % (r) 和 (l) 让横线向内收缩，视觉上分开左右两部分
%\midrule
%\rowcolor{grayhead}
\multicolumn{1}{c}{\textbf{Abbr.}} & \multicolumn{1}{c}{\textbf{Category}} & \multicolumn{1}{c}{\textbf{\# Contracts}} & \multicolumn{1}{c}{\textbf{\# Slices}} & 
\multicolumn{1}{c}{\textbf{Abbr.}} & \multicolumn{1}{c}{\textbf{Category}} & \multicolumn{1}{c}{\textbf{\# Contracts}} & \multicolumn{1}{c}{\textbf{\# Slices}} \\
\midrule

% 数据行 (注意左边只有5行，右边有6行，所以最后一行左边要空着)
\textbf{LIF} & Logic Implementation Flaws & 136 & 1,527 & \textbf{AC} & Access Control & 144 & 1,576 \\
\textbf{BPF} & Business Protocol Flaws & 120 & 1,236 & \textbf{DoS} & Denial of Service & 132 & 1,418 \\
\textbf{GAR} & Governance Authority Risks & 122 & 1,296 & \textbf{AE} & Arithmetic Errors & 142 & 1,484 \\
\textbf{LVD} & Liquidity Valuation Distortion & 136 & 1,574 & \textbf{RE} & Reentrancy & 146 & 1,449 \\
\textbf{TLS} & Transient Liquidity Shock & 136 & 1,511 & \textbf{TD} & Timestamp Dependency & 143 & 1,703 \\
% 最后一行：左边空，右边有数据
 & & & & \textbf{UC} & Unchecked Call & 133 & 1,380 \\

\bottomrule
\end{tabular}
\end{table}

\subsubsection*{\textbf{Data Preprocessing}}

Given that some samples collected in $I_{1(\text{core})}$ and $I_{1(\text{baseline})}$ only contain function snippets where the flaws are located and lack full source code, we use AI generation techniques to complete the full contract code. To ensure dataset accuracy and standardization, we strictly verify these completed contracts from both syntactic and semantic dimensions:

\begin{itemize}
\item \textbf{Compilability Verification}: We conduct comprehensive compilation tests on the completed contracts to exclude spurious or invalid code caused by Large Model Hallucination.

\item \textbf{Existence Verification}: We invite four graduate students experienced in blockchain security to conduct a comprehensive semantic audit of the contracts, ensuring that the flaws/vulnerabilities in the dataset genuinely exist and are exploitable.
\end{itemize}

After manual repair and screening, all contracts in the final dataset must pass the above dual verification, as detailed in Table~\ref{tab:dataset_qa_stats}.

% \definecolor{grayhead}{gray}{0.9} % 如果没定义颜色，请加上这一行

\begin{table}[htbp]
\centering
\caption{\textbf{Statistics of Verification Process and Contract Scale Distribution.} \textit{Note}: Contracts containing irreparable hallucinations or non-exploitable logic were regenerated.}
\label{tab:dataset_qa_stats}
\renewcommand{\arraystretch}{1} % 增加行高，更易阅读
\scriptsize % 使用稍大一点的字体 (scriptsize太小了，footnotesize比较合适)
\setlength{\tabcolsep}{8pt} % 适当增加列间距

% ==================== Panel A ====================
\begin{tabular}{@{} lrrr @{}}
\toprule
\rowcolor{grayhead}
\multicolumn{4}{c}{\textbf{Panel A: Statistics of Verification \& Rectification Process}} \\
\midrule
\multicolumn{1}{l}{\textbf{Verification Stage}} & 
\multicolumn{1}{r}{\textbf{Initial Candidates}} & 
\multicolumn{1}{r}{\textbf{Repaired}} & 
\multicolumn{1}{r}{\textbf{Final Retained}} \\
\midrule
AI-Based Completion        & 1,490 & -   & 1,490 \\
Compilability Verification & 1,490 & 290 & 1,490 \\
Existence Verification     & 1,490 & 30  & 1,490 \\
\bottomrule
\end{tabular}

\vspace{10pt} % 增加两个面板之间的间距

% ==================== Panel B ====================
\begin{tabular}{@{} l rr c rr @{}} % 中间加一个 c 列作为间隔，代替竖线
\toprule
\rowcolor{grayhead}
\multicolumn{6}{c}{\textbf{Panel B: Contract Size Distribution (Formatted SLOC)}} \\
\midrule
% 表头第一行
& \multicolumn{2}{c}{\textbf{$I_{1(\text{Core})}$}} & & \multicolumn{2}{c}{\textbf{$I_{1(\text{Baseline})}$}} \\
\cmidrule(lr){2-3} \cmidrule(lr){5-6} % 局部横线，强调分组，代替竖线的功能
% 表头第二行
\textbf{SLOC Range} & 
\multicolumn{1}{r}{\textbf{\# Contracts}} & 
\multicolumn{1}{r}{\textbf{Percentage}} & &
\multicolumn{1}{r}{\textbf{\# Contracts}} & 
\multicolumn{1}{r}{\textbf{Percentage}} \\
\midrule
Tiny ($< 50$)           & 105 & 16.2\% && 178 & 21.2\% \\
Small ($50 \sim 100$)   & 275 & 42.3\% && 329 & 39.2\% \\
Medium ($100 \sim 200$) & 223 & 34.3\% && 321 & 38.2\% \\
Large ($> 200$)         & 47  & 7.2\%  && 12  & 1.4\% \\
\midrule
\textbf{Total}          & \textbf{650} & \textbf{100\%} && \textbf{840} & \textbf{100\%} \\
\bottomrule
\end{tabular}
\end{table}

\subsection*{\textbf{Exp.1: Answer to RQ1.}}

\begin{figure*}[h!]
\centering
\includegraphics[width=\textwidth]{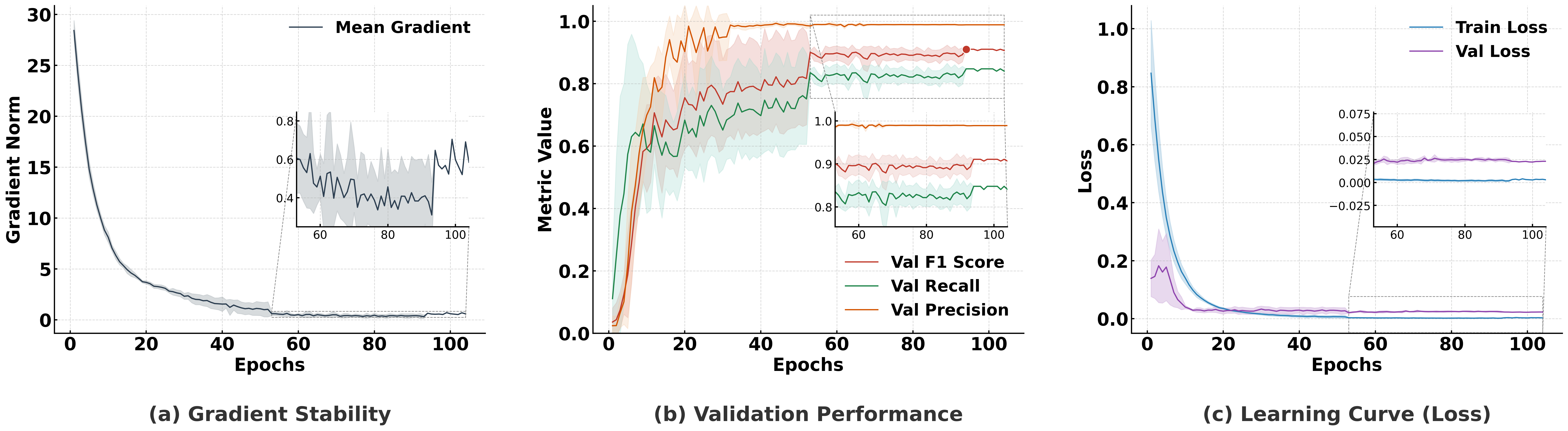}
\caption{\textbf{Performance dynamics of the DCN model during the $\mathbb{AIM}$ generation phase.} \textit{Note}: Shaded regions indicate the standard deviation across 5-fold cross-validation. In (a), the gradient spike near epoch 95 is a cross-validation artifact caused by a delayed fold triggering learning rate decay just prior to early stopping. In (b), the non-zero initial recall ($\approx 0.1$) stems from the positive sample weighting (\texttt{pos\_weight=6}) strategy, which forces the model to prioritize minority flaws from the onset.}
\label{fig: dcn}
\end{figure*}

Exp.1 aims to evaluate the effectiveness of the DCN model in extracting liquidity-related features from the $I_{1(\text{Core})}$ and to assess potential error propagation risks. To comprehensively gauge the model's convergence and generalization, we analyze the gradient dynamics, validation metrics, and loss curves, as shown in Fig.\ref{fig: dcn}.

\begin{enumerate}[label=(\arabic*), leftmargin=*, itemsep=3pt, topsep=3pt]

    \item \textbf{Training Dynamics and Convergence.} Subfigures (a) and (c) illustrate the training dynamics. The training gradient norm decreases significantly within the first 10 epochs and subsequently converges towards zero, indicating a clear and effective optimization direction. Simultaneously, the loss curves for both the training and validation sets decrease synchronously while maintaining a stable gap. This synchronization confirms that the model possesses strong generalization capabilities without exhibiting significant signs of overfitting.

    \item \textbf{Validation Performance Analysis.} Subfigure (b) details the performance on the validation set. The precision curve consistently hovers near 1.0, demonstrating the DCN model's exceptional ability to accurately filter out safe code snippets (reducing false positives). Conversely, the Recall and F1-score curves rise rapidly within the first 10 epochs but exhibit distinct behaviors before and after epoch 50:
    \begin{itemize}
        \item \textbf{Epochs 10--50}: The curves display significant volatility. This oscillation is attributed to the aggressive exploration of the optimization landscape driven by a relatively high initial learning rate, allowing the model to escape local optima but causing instability in metric evaluation.
        \item \textbf{Post-Epoch 50}: The fluctuations diminish significantly, and the metrics stabilize. This phenomenon indicates the successful activation of the learning rate decay scheduler, which reduced the step size to allow for fine-grained convergence into a robust local minimum.
    \end{itemize}

\end{enumerate}

\begin{tcolorbox}[
    breakable, % 允许盒子跨页断开
    enhanced, % 启用增强功能，如阴影
    colback=white, % 内部背景色为白色
    colframe=gray!50!black, % 边框颜色为灰色混合黑色
    arc=4mm, % 边框圆角半径
    boxrule=0.5pt, % 边框粗细
    left=3mm, right=3mm, top=3mm, bottom=3mm, % 内边距
    fonttitle=\itshape\bfseries, % 标题字体为斜体粗体
    title={Answer to RQ1.}, % 标题内容
    halign title=left, % 标题左对齐
    %valign title=top, % 标题顶部对齐
    title after break=\null, % 跨页后不显示标题
]
The DCN model excels in generating the $\mathbb{AIM}$, with stable training, consistent validation performance, and robust generalization.
\end{tcolorbox}

\subsection*{\textbf{Exp.2: Answer to RQ2.}}

Exp.2 aims to systematically evaluate the efficacy of the $\mathbb{AIM}$ component and the Four-Phase Collaborative Prompt System in enhancing LiquiLM's ability to audit liquidity flaws in contracts. Based on the dataset $I_{1(\text{Core})}$, we construct different variants of LiquiLM using Gemini 3 Pro and GPT-4o as backbone models, respectively, and conduct two sets of rigorous ablation studies. 

To comprehensively and fairly measure the model's performance in terms of reliability and robustness, we employ Precision, Recall, Specificity, and F1 Score as key evaluation metrics.

\begin{table*}[h!]
\centering
% 表格标题和 Note 保持不变
\caption{\textbf{Ablation Study: Reliability vs. Efficiency.} \textit{Note}: $\text{LiquiLM}_{\text{Gem/GPT}}$ employs a 3-round simulated verification in the Phase 2-1 feedback loop. Conversely, the \textit{w/o $\mathbb{AIM}$} baselines, lacking prior semantic guidance, default to Phase 2-2 and are limited to a single-pass verification judgment. Results reflect these specific strategies. Values are rounded ($\pm$0.01).}
\label{tab:full_metrics}

% 使用 resizebox 确保宽度自适应
\resizebox{\textwidth}{!}{
    % 增加行高，使表格更舒展
    \renewcommand{\arraystretch}{1}
    % 设置列间距
    \setlength{\tabcolsep}{5pt}
    
    \begin{tabular}{llcccccr} % 最后一列用 r (右对齐) 更符合数字习惯，前两列左对齐
        \toprule
        % 表头第一行
        \multirow{2}{*}{\textbf{Backbone}} & \multirow{2}{*}{\textbf{Method}} & \multicolumn{3}{c}{\textbf{Core Metrics (Macro)}} & \multicolumn{2}{c}{\textbf{F1 Score Analysis}} & \textbf{Efficiency} \\ 
        
        % 表头分隔线 (使用 cmidrule 并截断左右 lr)
        \cmidrule(lr){3-5} \cmidrule(lr){6-7} \cmidrule(lr){8-8}
        
        % 表头第二行
         & & \textbf{Precision} & \textbf{Recall} & \textbf{Specificity} & \textbf{Macro-F1} & \textbf{Weighted-F1} & \textbf{Avg. Tokens} \\
        \midrule
        
        % Gemini Group
        \multirow{2}{*}{\textbf{Gemini 3 pro}} 
          & $\text{LiquiLM}_{\text{Gem}}$ & 0.973 & \textbf{0.869} & 0.994 & \textbf{0.918} & \textbf{0.919} & 2.8k+ \\
          & \textit{w/o $\mathbb{AIM}$}     & 0.571 & 0.437 & 0.877 & 0.396 & 0.385 & 3.9k+ \\
        \midrule % 组间分隔线
        
        % GPT Group
        \multirow{2}{*}{\textbf{GPT-4o}} 
          & $\text{LiquiLM}_{\text{GPT}}$ & 0.976 & 0.846 & 0.995 & 0.906 & 0.907 & \textbf{2.6k+} \\
          & \textit{w/o $\mathbb{AIM}$}     & 0.598 & 0.451 & 0.873 & 0.420 & 0.407 & 6.3k+ \\
        \midrule
        
        % Bottom Group (AIM Only)
        \multicolumn{2}{c}{\textit{only $\mathbb{AIM}$}} & \textbf{0.995} & 0.764 & \textbf{0.999} & 0.856 & 0.861 & -- \\
        \bottomrule
    \end{tabular}
}
\end{table*}

Table~\ref{tab:full_metrics} details the results of the ablation experiments, clearly elucidating the contribution of each core component to LiquiLM's overall performance. Considering the imbalance in the distribution of flaw categories and the varying importance of different samples, we calculate and report both Macro-F1 and Weighted-F1. Overall, regardless of whether it is based on Gemini 3 Pro or GPT-4o, the fully configured LiquiLM framework achieves significant performance advantages in Recall, Specificity, and both F1 scores.

Based on the aforementioned experimental data and phenomenon analysis, we draw the following conclusions:

\begin{enumerate}[label=(\arabic*), leftmargin=*, itemsep=3pt, topsep=3pt]
\item \textbf{Contextual Dilemma of LLMs (\textit{w/o $\mathbb{AIM}$})}: Removing the Prior Knowledge provided by the list leads to significant performance degradation. This indicates that relying solely on raw source code forces LLMs to perform unguided blind retrieval within a vast semantic search space. Especially when facing contracts involving lengthy logical chains, LLMs confront both a scarcity of key context and an overload of redundant information. Consequently, they struggle to capture the semantic gap, leading to a substantial increase in missed flaws and computational overhead (e.g., token consumption surges by $\sim$140\% on GPT-4o). % [新增] 强调了无效剪枝导致的Token激增（基于6.3k vs 2.6k的数据）

\item \textbf{Semantic Understanding Limitations of Standalone DCN (\textit{only $\mathbb{AIM}$})}: Although this configuration performs excellently in precision (approaching 99.5\%), it fails to reach an ideal level in recall (only 76.4\%). This result reveals the dual nature of the DCN: it effectively filters out obvious safe code snippets and greatly suppresses false positives with zero LLM token consumption; % [新增] 强调了AIM作为过滤器的零成本特性
however, due to limited reasoning capabilities, it struggles with concealed flaws. This validates our design philosophy: the DCN should serve as a high-confidence, cost-efficient "pre-screening filter" rather than an independent endpoint.

\item \textbf{Necessity of the Four-Phase Collaborative Prompt System}: Experimental data show that when the LiquiLM framework is fully enabled, the model's recall increases over 8\% compared to using only $\mathbb{AIM}$, while precision shows only a marginal drop. This trade-off indicates that the four-stage synergistic mechanism successfully activates the dynamic reasoning capability of the LLM. It effectively captures deep logical flaws while maintaining high economic efficiency (avg. $\sim$2.7k tokens per contract), thereby achieving the overall optimal cost-performance ratio.

\item \textbf{Generalizability of $\mathbb{AIM}$}: Although Gemini 3 Pro performs slightly worse than GPT-4o in the unassisted environment, LiquiLM$_{Gem}$ integrated with $\mathbb{AIM}$ achieves a performance leap, with precision increasing by up to 40.2\% and recall by 43.2\%. This result demonstrates the model-agnostic nature of $\mathbb{AIM}$: by providing standardized feature guidance, it maximizes the reasoning potential of different backbones and bridges the capability gap, while significantly reducing token waste caused by redundant context processing.
\end{enumerate}

%Furthermore, we perform an in-depth analysis of the model's stability by comprehensively reviewing the audit reports generated by LiquiLM.

%\subsubsection*{\textbf{Reliability.}}

\begin{figure*}
\centering
\includegraphics[width=\textwidth]{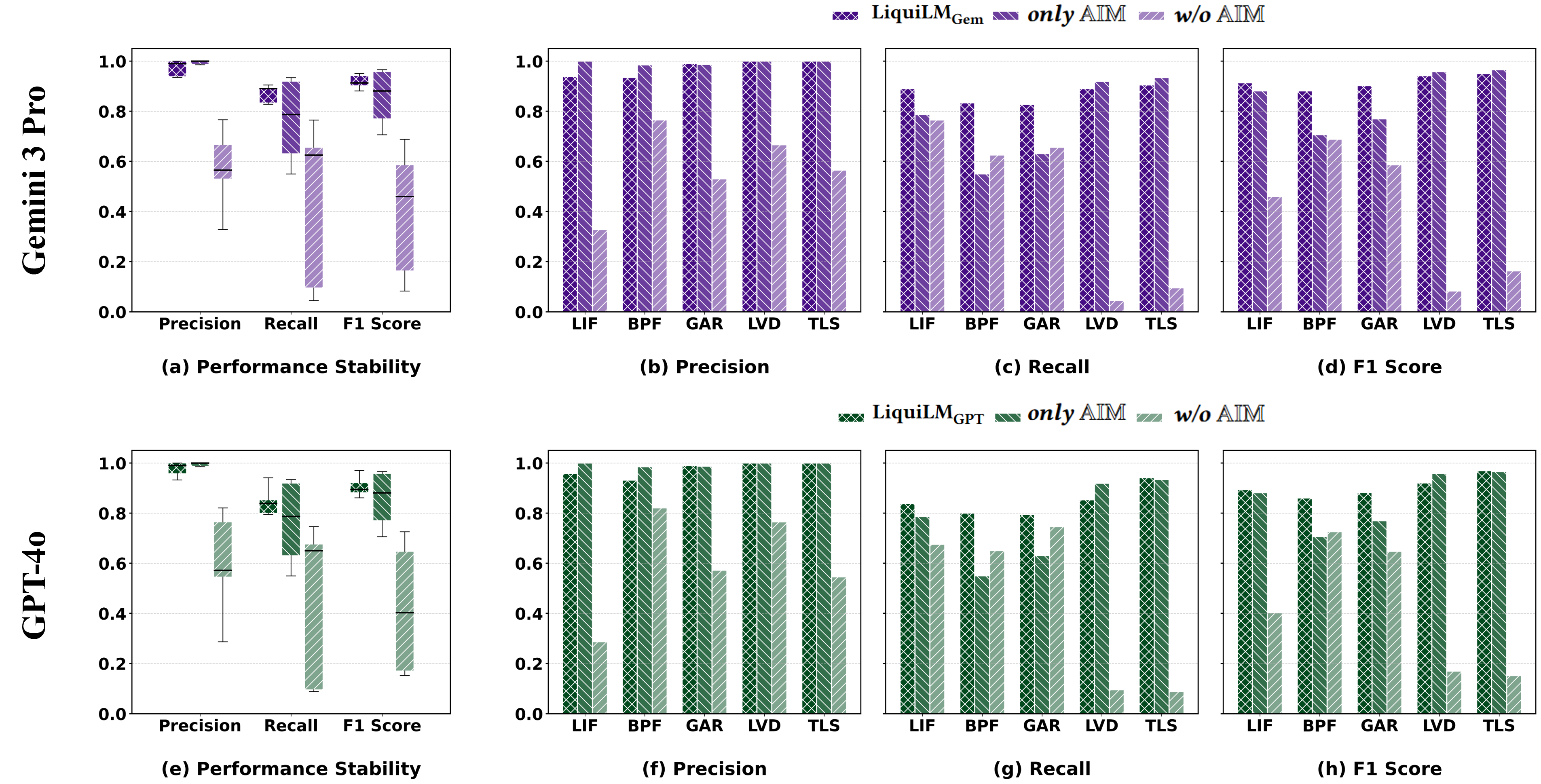}
\caption{\textbf{Fine-grained reliability evaluation across five liquidity flaw types.} \textit{Note}: Subfigures (a) and (e) display the distribution box plots of the three metrics; Subfigures (b)-(d) and (f)-(h) detail the specific performance of Precision, Recall, and F1 Score across different flaw categories, respectively.}
\label{fig: comparison_of_bar}
\end{figure*}

To further dissect the performance mechanisms underlying the macroscopic metrics, Fig.~\ref{fig: comparison_of_bar} presents the fine-grained evaluation results for five specific flaw types, wherein the subfigures reaffirm the full framework's advantage in overall reliability. 

Notably, the recall subfigures (c) and (g) indicate that while relying solely on contract source code (\textit{w/o $\mathbb{AIM}$}) suffices to cover most endogenous flaws (e.g., LIF, GAR), detection capability suffers a precipitous decline when confronting exogenous flaws (LVD and TLS) characterized by higher logical concealment or complex state constraints. This quantitative evidence confirms that LLMs require the feature guidance provided by $\mathbb{AIM}$ to focus the model's attention, enabling the capture of deep semantics and thereby bridging the semantic gap between low-level code implementation and high-level intent.

\begin{tcolorbox}[
    breakable, % 允许盒子跨页断开
    enhanced, % 启用增强功能，如阴影
    colback=white, % 内部背景色为白色
    colframe=gray!50!black, % 边框颜色为灰色混合黑色
    arc=4mm, % 边框圆角半径
    boxrule=0.5pt, % 边框粗细
    left=3mm, right=3mm, top=3mm, bottom=3mm, % 内边距
    fonttitle=\itshape\bfseries, % 标题字体为斜体粗体
    title={Answer to RQ2.}, % 标题内容
    halign title=left, % 标题左对齐
    %valign title=top, % 标题顶部对齐
    title after break=\null, % 跨页后不显示标题
]
The ablation studies demonstrate LiquiLM's robust performance in detecting liquidity flaws. This superiority stems primarily from the synergistic effect between $\mathbb{AIM}$ and the four-stage collaborative prompting system; the absence of either component significantly compromises the model's reliability.
\end{tcolorbox}

\subsection*{\textbf{Exp.3: Answer to RQ3.}}

Exp.3 benchmarks LiquiLM against state-of-the-art smart contract vulnerability detection tools. To ensure broad and representative coverage, we select four traditional analysis tools (Slither \cite{feist2019slither}, Oyente+ \cite{luu2016making}, Aderyn \cite{aderyn}, and Solhint \cite{solhint}) and two fine-tuned LLM-based tools (GPTLens \cite{hu2023large} and Smart-LLaMA-DPO \cite{yu2025smart}) as baselines.

Given that these baseline methods focus on specific dimensions distinct from the liquidity flaws defined in this study, we conduct the evaluation on the $I_{1(\text{Baseline})}$ dataset. This dataset contains standard vulnerabilities that are well-adapted to the capabilities of the baseline tools, ensuring a fair comparison.

\begin{table}[h!]
\centering
\caption{\textbf{Performance Comparison of $\text{LiquiLM}$ vs. Baselines (Part 1)}. \textit{Note}: Evaluation is limited to contracts yielding valid tool outputs. GPTLens utilizes Gemini 3 Pro, while Smart-LLaMA-DPO uses a fine-tuned LLaMA3.1-8B. Smart-LLaMA-DPO metrics may be overestimated due to its focus on specific vulnerability detection rather than full auditing. Values are rounded ($\pm$0.01).}
\label{tab:vulnerability_comparison1}
\renewcommand{\arraystretch}{1.2}
\small
% 注意：由于横向内容较多，建议使用 textwidth。如果是双栏排版，建议将 table 改为 table* 环境
\resizebox{\textwidth}{!}{ 
\begin{tabular}{c|ccc|ccc|ccc} 
    \toprule
    \multirow{2}{*}{\textbf{Methods}} & 
    \multicolumn{3}{c|}{\textbf{AC}} & 
    \multicolumn{3}{c|}{\textbf{DoS}} & 
    \multicolumn{3}{c}{\textbf{AE}} \\ 
    \cline{2-10}
     & \textbf{P(\%)} & \textbf{R(\%)} & \textbf{F1(\%)} & \textbf{P(\%)} & \textbf{R(\%)} & \textbf{F1(\%)} & \textbf{P(\%)} & \textbf{R(\%)} & \textbf{F1(\%)} \\ 
    \midrule
    
    % 基线模型数据 (请替换为实际数据)
    \textbf{Aderyn}      & 0.189 & 0.868 & 0.310 & 0.334 & 0.962 & 0.496 & 0.273 & 0.021 & 0.039 \\
    \textbf{Oyente+}     & 0.000 & 0.000 & 0.000 & 0.117 & 0.349 & 0.175 & 0.229 & 0.923 & 0.337 \\
    \textbf{Slither}     & 0.151 & 0.569 & 0.238 & 0.376 & 0.818 & 0.516 & 0.150 & 0.106 & 0.125 \\
    \textbf{Solhint}     & 0.195 & 0.993 & 0.325 & 0.000 & 0.000 & 0.000 &  N/A  &  N/A  &  N/A  \\
    \hline
    
    % GPT4scv
%    \textbf{GPTLens}            & 0.210 & 0.951 & 0.343 & 0.229 & 0.788 & 0.354 & 0.668 & 0.929 & 0.777 \\
    \textbf{GPTLens}            & 0.209 & 0.972 & 0.344 & 0.232 & 0.780 & 0.357 & 0.625 & 0.922 & 0.745 \\
    \textbf{Smart-LLaMA-DPO}    & 1.000 & 0.4476 & 0.618 &  N/A  &  N/A  &  N/A  & 0.000 & 0.000 & 0.000 \\
    
    % 我的模型
%    \rowcolor{lightblue}
%    \textbf{$\text{LiquiLM}_{\text{GPT}}$} & 
%    \textbf{00.00} & \textbf{00.00} & \textbf{00.00} & 
%    \textbf{00.00} & \textbf{00.00} & \textbf{00.00} & 
%    \textbf{00.00} & \textbf{00.00} & \textbf{00.00} \\
    \rowcolor{lightblue}
    \textbf{$\text{LiquiLM}_{\text{Gem}}$} & 
    \textbf{0.734} & \textbf{0.965} & \textbf{0.834} & 
    \textbf{0.935} & \textbf{0.977} & \textbf{0.956} & 
    \textbf{0.593} & \textbf{0.950} & \textbf{0.730} \\
    \bottomrule
\end{tabular}}
\end{table}

\begin{table}[h!]
\centering
\caption{\textbf{Performance Comparison of $\text{LiquiLM}$ vs. Baselines (Part 2)}. \textit{Note}: Evaluation is limited to contracts yielding valid tool outputs. GPTLens utilizes Gemini 3 Pro, while Smart-LLaMA-DPO uses a fine-tuned LLaMA3.1-8B. Smart-LLaMA-DPO metrics may be overestimated due to its focus on specific vulnerability detection rather than full auditing. Values are rounded ($\pm$0.01).}
\label{tab:vulnerability_comparison2}
\renewcommand{\arraystretch}{1.2}
\small
% 注意：由于横向内容较多，建议使用 textwidth。如果是双栏排版，建议将 table 改为 table* 环境
\resizebox{\textwidth}{!}{ 
\begin{tabular}{c|ccc|ccc|ccc} 
    \toprule
    \multirow{2}{*}{\textbf{Methods}} & 
    \multicolumn{3}{c|}{\textbf{RE}} & 
    \multicolumn{3}{c|}{\textbf{TD}} & 
    \multicolumn{3}{c}{\textbf{UC}} \\ 
    \cline{2-10}
     & \textbf{P(\%)} & \textbf{R(\%)} & \textbf{F1(\%)} & \textbf{P(\%)} & \textbf{R(\%)} & \textbf{F1(\%)} & \textbf{P(\%)} & \textbf{R(\%)} & \textbf{F1(\%)} \\ 
    \midrule
    
    % 基线模型数据 (请替换为实际数据)
    \textbf{Aderyn}      & 0.483 & 0.993 & 0.650 & 0.600 & 0.021 & 0.041 & 0.274 & 0.895 & 0.420 \\
    \textbf{Oyente+}     & 0.727 & 0.658 & 0.691 & 0.154 & 0.014 & 0.026 &  N/A  &  N/A  &  N/A  \\
    \textbf{Slither}     & 0.305 & 1.000 & 0.468 & 0.418 & 0.909 & 0.573 & 0.541 & 0.850 & 0.661\\
    \textbf{Solhint}     & 0.692 & 0.445 & 0.542 & 0.000 & 0.000 & 0.000 & 0.627 & 1.000 & 0.771 \\
    \hline
    
    % GPT4scv
%    \textbf{GPTLens}            & 0.806 & 0.966 & 0.879 & 0.583 & 0.245 & 0.345 & 0.524 & 0.729 & 0.610 \\
    \textbf{GPTLens}            & 0.755 & 0.993 & 0.858 & 0.662 & 0.315 & 0.427 & 0.524 & 0.744 & 0.615 \\
    \textbf{Smart-LLaMA-DPO}    & 1.000 & 0.048 & 0.091 & 1.000 & 0.965 & 0.982 & 1.000 & 0.068 & 0.127 \\
    
    % 我的模型
%    \rowcolor{lightblue}
%    \textbf{$\text{LiquiLM}_{\text{GPT}}$} & 
%    \textbf{00.00} & \textbf{00.00} & \textbf{00.00} & 
%    \textbf{00.00} & \textbf{00.00} & \textbf{00.00} & 
%    \textbf{00.00} & \textbf{00.00} & \textbf{00.00} \\
    \rowcolor{lightblue}
    \textbf{$\text{LiquiLM}_{\text{Gem}}$} & 
    \textbf{0.899} & \textbf{0.980} & \textbf{0.938} & 
    \textbf{0.978} & \textbf{0.923} & \textbf{0.950} & 
    \textbf{0.942} & \textbf{0.970} & \textbf{0.956} \\
    \bottomrule
\end{tabular}}
\end{table}

The performance comparison between LiquiLM and baseline tools across six traditional vulnerability categories is presented in Table~\ref{tab:vulnerability_comparison1} and Table~\ref{tab:vulnerability_comparison2}. Based on the experimental data, we summarize the following key observations:

\begin{enumerate}[label=(\arabic*), leftmargin=*, itemsep=3pt, topsep=3pt] 

\item \textbf{Achieving a balance between precision and recall}: 
Traditional analysis tools generally exhibit extremely high false positive rates on these vulnerabilities (e.g., Slither shows 100\% recall on RE detection but only 30.5\% precision). In contrast, LiquiLM introduces a multi-stage verification mechanism, which substantially filters out false positive noise while maintaining high recall, significantly improving detection precision (e.g., achieving an F1-score of 93.8\% on RE).

\item \textbf{Superior generalization capability over fine-tuned models}: 
Although domain-specific fine-tuned models achieve peak performance on tasks they are trained on (e.g., Smart-LLaMA-DPO reaches an optimal F1 of 98.2\% on TD), they are prone to overfitting and lack generalization capabilities for unseen tasks (e.g., suffering from severe false negatives on AE). LiquiLM exhibits more balanced robustness and adapts effectively to various types of vulnerability detection tasks.

\item \textbf{Advantage in complex semantic scenarios}: 
On vulnerability types that rely heavily on business logic and semantic understanding (such as DoS and UC), LiquiLM performs best, with F1-scores reaching 95.6\% in both cases, significantly outperforming existing SOTA tools (Slither's 51.6\% on DoS and Solhint's 77.1\% on UC). This indicates that LiquiLM effectively overcomes the limitations of traditional static analysis tools that rely solely on shallow pattern matching, enabling the identification of deep logical flaws via the semantic context provided by the DCN.

\item \textbf{Limitations on syntax-based vulnerabilities}: 
We observe that compared to the auditing performance on liquidity flaws (Exp. 1), LiquiLM's precision drops notably on vulnerabilities leaning towards syntactic features (e.g., only 59.3\% precision on AE), falling short of the general framework GPTLens (62.5\%). This is mainly due to the design trade-off of the DCN module: The DCN aims to capture deep liquidity logic through complex semantic interactions; however, when handling traditional vulnerabilities that require only simple syntactic analysis, this deep semantic mapping potentially introduces unnecessary semantic noise, thereby interfering with the downstream LLM's judgment.

\end{enumerate}

\begin{tcolorbox}[
    breakable, % 允许盒子跨页断开
    enhanced, % 启用增强功能，如阴影
    colback=white, % 内部背景色为白色
    colframe=gray!50!black, % 边框颜色为灰色混合黑色
    arc=4mm, % 边框圆角半径
    boxrule=0.5pt, % 边框粗细
    left=3mm, right=3mm, top=3mm, bottom=3mm, % 内边距
    fonttitle=\itshape\bfseries, % 标题字体为斜体粗体
    title={Answer to RQ3.}, % 标题内容
    halign title=left, % 标题左对齐
    %valign title=top, % 标题顶部对齐
    title after break=\null, % 跨页后不显示标题
]
The results of Exp.3 demonstrate LiquiLM's overall superiority over baselines in detecting traditional vulnerabilities, although the DCN module's focus on deep semantic interactions introduces noise that slightly compromises precision on simpler, syntax-based tasks.
\end{tcolorbox}

\subsection*{\textbf{Exp.4: Answer to RQ4.}}

Exp.4 is designed to validate the effectiveness of LiquiLM in auditing liquidity flaws within non-synthetic and uncontrolled real-world scenarios.

We input all 1,380 contracts from $I_{1(\text{Real})}$ into $\text{LiquiLM}_{\text{Gem}}$ for comprehensive scanning and analysis. The audit report reveals that $\text{LiquiLM}_{\text{Gem}}$ successfully identifies potential liquidity security risks in 238 contracts (corresponding to an alert rate of approximately 17.2\%). Following manual expert review and responsible disclosure to project teams, we successfully obtained 10 CVE (Common Vulnerabilities and Exposures) certifications.

\begin{figure*}
\centering\includegraphics[width=\textwidth]{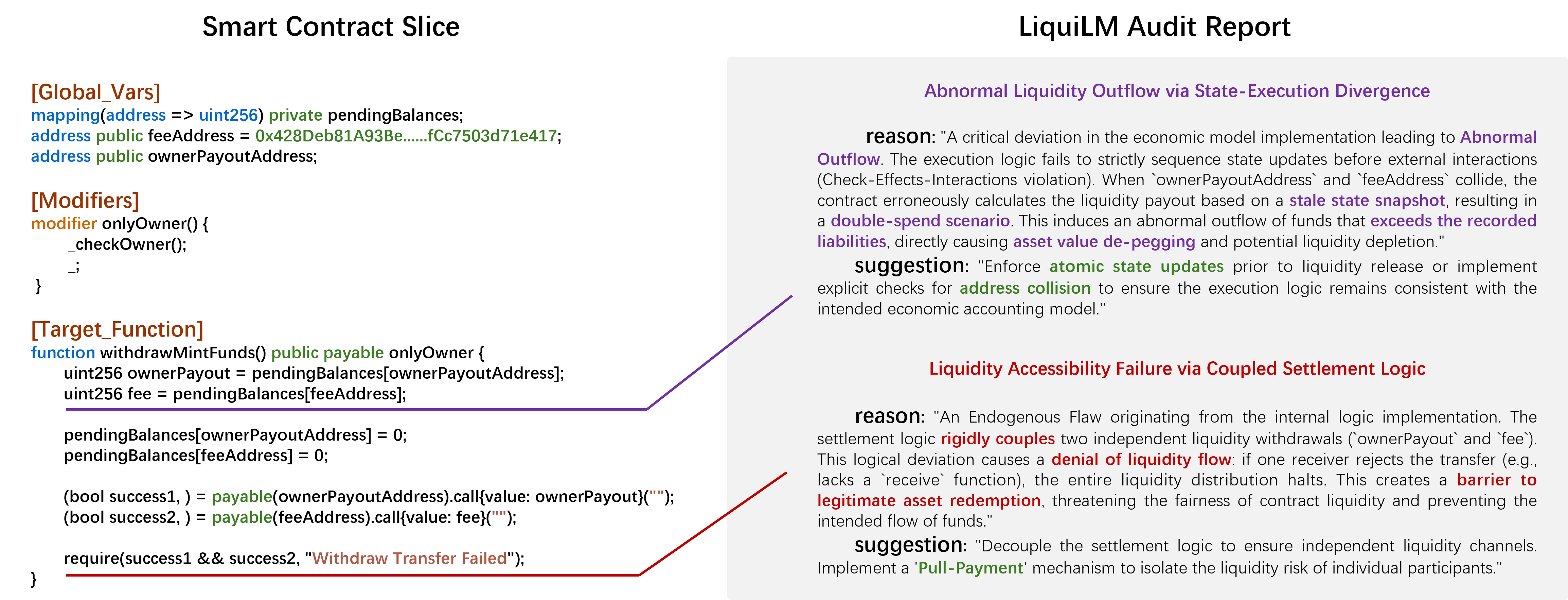}
\caption{\textbf{Example of LiquiLM Audit Results.} \textit{Note}: For clarity, we condense the audit report content, retaining only the "reason" and "suggestion" fields from the original report.}
\label{fig: audit}
\end{figure*}

Fig.~\ref{fig: audit} presents the audit results generated by LiquiLM for a contract slice from $I_{1(\text{Real})}$, which contains two distinct LIFs.

\begin{tcolorbox}[
    breakable, % 允许盒子跨页断开
    enhanced, % 启用增强功能，如阴影
    colback=white, % 内部背景色为白色
    colframe=gray!50!black, % 边框颜色为灰色混合黑色
    arc=4mm, % 边框圆角半径
    boxrule=0.5pt, % 边框粗细
    left=3mm, right=3mm, top=3mm, bottom=3mm, % 内边距
    fonttitle=\itshape\bfseries, % 标题字体为斜体粗体
    title={Answer to RQ4.}, % 标题内容
    halign title=left, % 标题左对齐
    %valign title=top, % 标题顶部对齐
    title after break=\null, % 跨页后不显示标题
]
Exp.4 confirms that LiquiLM possesses industrial-grade flaw mining capabilities, demonstrating its practical value and credibility in countering complex and dynamic liquidity manipulation threats in the real world.
\end{tcolorbox}
\section{\textbf{Related Work}}

\subsubsection*{\textbf{Traditional Flaw Detection Methods}}

Currently, traditional detection methods have become quite mature, including static analysis tools \cite{schneidewind2020ethor,xue2020cross}; dynamic analysis tools \cite{eshghie2021dynamic,ding2020function,li2022smartfast}; formal verification tools \cite{baba2024modeling,wang2020formal}; and systematic penetration testing approaches adapted for comprehensive system security \cite{zhang2025penetration}. These methods perform well in detecting common types of flaws and can effectively reduce explicit flaws in the early stages of development. By learning from these methods, we further improved the design and implementation of our suspicion list generation module, making the identification of flawed items more comprehensive and accurate.

\subsubsection*{\textbf{LLMs-Supported Auditing Methods}}

Hu et al. introduce GPTL ENS, an adversarial framework where LLMs serve as AUDITOR for comprehensive contract flaw detection and CRITIC to reduce false positives \cite{hu2023large}. Additionally, numerous studies adopt LLMs as the primary method for smart contract auditing \cite{ince2024detect,xia2024auditgpt}.

\subsubsection*{\textbf{LiquiLM Framework}}

Compared to LLM-supported auditing methods, traditional flaw detection faces challenges like higher costs, limited portability, and path explosion in complex contracts, but offers more reproducible results and higher accuracy for common flaws. LiquiLM integrates LLMs with DCN, combining the rigor of traditional auditing with the efficiency of modern techniques. The Audit-Informed Manifest enhances the contextual information for LLMs, enabling more precise analysis, while the Four-Phase Collaborative Prompt System improves audit report accuracy and reliability. To ensure versatility and scalability, LiquiLM adopts a general framework for flaw detection, facilitating integration into various blockchain scenarios and supporting future applications across diverse domains.
\section{Conclusion}

In this paper, we propose LiquiLM, a framework that innovatively integrates LLMs with DCN, aiming to achieve precise auditing and deep explanation of smart contract liquidity flaws. Leveraging DCN, we establish deep semantic correlations between target contracts and the liquidity flaw semantic corpus, thereby capturing the hidden semantic features of liquidity flaws. Based on this, the method generates a comprehensive Audit-Informed Manifest, providing high-value contextual prior information for subsequent analysis. Furthermore, to maximize auditing efficacy, we design a Four-Phase Collaborative Prompting System. Compared with traditional single-stage prompting, this system significantly enhances the reliability and stability of final audit reports through hierarchical progressive analysis and real-time verification mechanisms. Comprehensive evaluation experiments demonstrate that LiquiLM exhibits superior scenario adaptability and significant flaw coverage advantages in auditing and explaining smart contract liquidity flaws.

Future work will focus on the following aspects: first, further expanding the experimental dataset to cover more diverse contract scenarios; second, conducting targeted model fine-tuning to enhance domain knowledge; and finally, continuously optimizing the auditing workflow of LiquiLM to comprehensively improve the accuracy and generalization capability of liquidity flaw detection.

%\section{Data Availability}

%To ensure the reproducibility of this work, we ensure that the fundamental resources are accessible, including:

%\begin{itemize}
%    \item All datasets extracted from the literature (including previously published work by the authors) are publicly available.
    
%    \item All necessary code, dependency version information, and instructions required for the experiments are openly provided to ensure consistency with the paper’s results.
    
%    \item All experiments are conducted on a computing node equipped with an NVIDIA RTX 4060 Ti GPU, and we specify details such as the operating system version, CUDA version, framework version, and hyperparameter settings.
%\end{itemize}

%\noindent The code and experimental data for LiquiLM have been made publicly available at the following anonymous link: \url{https://figshare.com/s/3bfe12f83fa35ed301c3}.

%To ensure reproducibility, we provide publicly available datasets from existing literature (which may include previously published work by the authors), openly shared code with dependency versions and experimental instructions, and experiment details conducted on a node with an NVIDIA RTX 4060 Ti GPU, including operating system, CUDA version, framework version, and hyperparameter settings.

\bibliographystyle{ACM-Reference-Format}
\bibliography{references}

@String{Computing = "Computing" }

@String{Computer = "{IEEE} Computer" }

@String{Springer = "Springer-Verlag" }

@article{abgaryan2024proof,
  title={Proof of Efficient Liquidity: A Staking Mechanism for Capital Efficient Liquidity},
  author={Abgaryan, Arman and Sharma, Utkarsh and Tobkin, Joshua},
  journal={arXiv preprint arXiv:2401.04521},
  year={2024}
}

@inproceedings{grishchenko2018foundations,
  title={Foundations and tools for the static analysis of ethereum smart contracts},
  author={Grishchenko, Ilya and Maffei, Matteo and Schneidewind, Clara},
  booktitle={Proceedings of the 30th International Computer Aided Verification (CAV)},
  pages={51--78},
  year={2018},
}

@article{ji2023effuzz,
  title={Effuzz: Efficient fuzzing by directed search for smart contracts},
  author={Ji, Songyan and Wu, Jin and Qiu, Junfu and Dong, Jian},
  journal={Information and Software Technology},
  volume={159},
  pages={107213--107225},
  year={2023},
}

@inproceedings{pani2023smartfuzzdrivergen,
  title={Smartfuzzdrivergen: Smart contract fuzzing automation for golang},
  author={Pani, Siddhasagar and Nallagonda, Harshita Vani and Vigneswaran and Medicherla, Raveendra Kumar and Rajan, M},
  booktitle={Proceedings of the 16th Innovations in Software Engineering Conference (ISEC)},
  pages={1--11},
  year={2023}
}

@article{almakhour2020verification,
  title={Verification of smart contracts: A survey},
  author={Almakhour, Mouhamad and Sliman, Layth and Samhat, Abed Ellatif and Mellouk, Abdelhamid},
  journal={Pervasive and Mobile Computing},
  volume={67},
  pages={101227--101246},
  year={2020},
}

@article{boi2024smart,
  title={Smart Contract Vulnerability Detection: The Role of Large Language Model (LLM)},
  author={Boi, Biagio and Esposito, Christian and Lee, Sokjoon},
  journal={ACM SIGAPP Applied Computing Review},
  volume={24},
  number={2},
  pages={19--29},
  year={2024},
}

@inproceedings{luo2024fellmvp,
  title={FELLMVP: An Ensemble LLM Framework for Classifying Smart Contract Vulnerabilities},
  author={Luo, Yu and Xu, Weifeng and Andersson, Karl and Hossain, Mohammad Shahadat and Xu, Dianxiang},
  booktitle={Proceedings of the IEEE International Conference on Blockchain (ICBC)},
  pages={89--96},
  year={2024},
}

@inproceedings{boi2024vulnhunt,
  title={VulnHunt-GPT: a Smart Contract vulnerabilities detector based on OpenAI chatGPT},
  author={Boi, Biagio and Esposito, Christian and Lee, Sokjoon},
  booktitle={Proceedings of the 39th ACM/SIGAPP Symposium on Applied Computing (SAC)},
  pages={1517--1524},
  year={2024}
}

@article{chen2023chatgpt,
  title={When chatgpt meets smart contract vulnerability detection: How far are we?},
  author={Chen, Chong and Su, Jianzhong and Chen, Jiachi and Wang, Yanlin and Bi, Tingting and Yu, Jianxing and et al.},
  journal={arXiv preprint arXiv:2309.05520},
  year={2023}
}

@inproceedings{ding2024evaluation,
  title={Evaluation of Question-Answering Based Text Summarization using LLM Invited Paper},
  author={Ding, Junhua and Nguyen, Huyen and Chen, Haihua},
  booktitle={Proceedings of the IEEE International Conference on Artificial Intelligence Testing (AITest)},
  pages={142--149},
  year={2024},
}

@inproceedings{huang2023towards,
  title={Towards making the most of llm for translation quality estimation},
  author={Huang, Hui and Wu, Shuangzhi and Liang, Xinnian and Wang, Bing and Shi, Yanrui and Wu, Peihao and Yang, Muyun and Zhao, Tiejun},
  booktitle={Proceedings of the CCF International Conference on Natural Language Processing and Chinese Computing (NLPCC)},
  pages={375--386},
  year={2023},
}

@inproceedings{nam2024using,
  title={Using an llm to help with code understanding},
  author={Nam, Daye and Macvean, Andrew and Hellendoorn, Vincent and Vasilescu, Bogdan and Myers, Brad},
  booktitle={Proceedings of the IEEE/ACM 46th International Conference on Software Engineering (ICSE)},
  pages={1--13},
  year={2024}
}

@article{huang2023bias,
  title={Bias assessment and mitigation in llm-based code generation},
  author={Huang, Dong and Bu, Qingwen and Zhang, Jie and Xie, Xiaofei and Chen, Junjie and Cui, Heming},
  journal={arXiv preprint arXiv:2309.14345},
  year={2023}
}

@article{deng2024wav2prompt,
  title={Wav2Prompt: End-to-End Speech Prompt Generation and Tuning For LLM in Zero and Few-shot Learning},
  author={Deng, Keqi and Sun, Guangzhi and Woodland, Philip C},
  journal={arXiv preprint arXiv:2406.00522},
  year={2024}
}

@article{john2023smart,
  title={Smart contracts and decentralized finance},
  author={John, Kose and Kogan, Leonid and Saleh, Fahad},
  journal={Annual Review of Financial Economics},
  volume={15},
  number={1},
  pages={523--542},
  year={2023},
}

@inproceedings{gunjal2024detecting,
  title={Detecting and preventing hallucinations in large vision language models},
  author={Gunjal, Anisha and Yin, Jihan and Bas, Erhan},
  booktitle={Proceedings of the AAAI Conference on Artificial Intelligence (AAAI)},
  volume={38},
  number={16},
  pages={18135--18143},
  year={2024}
}

@inproceedings{schneidewind2020ethor,
  title={ethor: Practical and provably sound static analysis of ethereum smart contracts},
  author={Schneidewind, Clara and Grishchenko, Ilya and Scherer, Markus and Maffei, Matteo},
  booktitle={Proceedings of the ACM SIGSAC Conference on Computer and Communications Security (CCS)},
  pages={621--640},
  year={2020}
}

@inproceedings{xue2020cross,
  title={Cross-contract static analysis for detecting practical reentrancy vulnerabilities in smart contracts},
  author={Xue, Yinxing and Ma, Mingliang and Lin, Yun and Sui, Yulei and Ye, Jiaming and Peng, Tianyong},
  booktitle={Proceedings of the 35th IEEE/ACM International Conference on Automated Software Engineering (ASE)},
  pages={1029--1040},
  year={2020}
}

@inproceedings{eshghie2021dynamic,
  title={Dynamic vulnerability detection on smart contracts using machine learning},
  author={Eshghie, Mojtaba and Artho, Cyrille and Gurov, Dilian},
  booktitle={Proceedings of the 25th International Conference on Evaluation and Assessment in Software Engineering (EASE)},
  pages={305--312},
  year={2021}
}

@article{ding2020function,
  title={Function-level dynamic monitoring and analysis system for smart contract},
  author={Ding, Yi and Wang, Chenshuo and Zhong, Qionghui and Li, Haisheng and Tan, Jinjing and Li, Jie},
  journal={IEEE Access},
  volume={8},
  pages={229161--229172},
  year={2020},
}

@article{li2022smartfast,
  title={SmartFast: an accurate and robust formal analysis tool for Ethereum smart contracts},
  author={Li, Zhaoxuan and Lu, Siqi and Zhang, Rui and Xue, Rui and Ma, Wenqiu and Liang, Rujin and et al.},
  journal={Empirical Software Engineering},
  volume={27},
  number={7},
  pages={197},
  year={2022},
}

@inproceedings{wang2020formal,
  title={A formal verification method for smart contract},
  author={Wang, Xiaobing and Yang, Xiaoyu and Li, Chunyi},
  booktitle={Proceedings of the 7th International Conference on Dependable Systems and Their Applications (DSA)},
  pages={31--36},
  year={2020},
}

@inproceedings{ince2024detect,
  title={Detect Llama-Finding Vulnerabilities in Smart Contracts Using Large Language Models},
  author={Ince, Peter and Luo, Xiapu and Yu, Jiangshan and Liu, Joseph K and Du, Xiaoning},
  booktitle={Proceedings of the Australasian Conference on Information Security and Privacy (ACISP)},
  pages={424--443},
  year={2024},
}

@article{xia2024auditgpt,
  title={AuditGPT: Auditing Smart Contracts with ChatGPT},
  author={Xia, Shihao and Shao, Shuai and He, Mengting and Yu, Tingting and Song, Linhai and Zhang, Yiying},
  journal={arXiv preprint arXiv:2404.04306},
  year={2024}
}

@inproceedings{baba2024modeling,
  title={Modeling and verification of solidity smart contracts with the B method},
  author={Baba, Fay{\c{c}}al and Mammar, Amel and Frappier, Marc and Laleau, R{\'e}gine},
  booktitle={Proceedings of the 28th International Conference on Engineering of Complex Computer Systems (ICECCS)},
  pages={159--178},
  year={2024}
}

@article{yu2025smart,
  title={Smart-LLaMA-DPO: Reinforced Large Language Model for Explainable Smart Contract Vulnerability Detection},
  author={Yu, Lei and Huang, Zhirong and Yuan, Hang and Cheng, Shiqi and Yang, Li and Zhang, Fengjun and Shen, Chenjie and Ma, Jiajia and Zhang, Jingyuan and Lu, Junyi and others},
  journal={Proceedings of the ACM on Software Engineering},
  volume={2},
  number={ISSTA},
  pages={182--205},
  year={2025},
  publisher={ACM New York, NY, USA}
}

@inproceedings{feist2019slither,
  title={Slither: a static analysis framework for smart contracts},
  author={Feist, Josselin and Grieco, Gustavo and Groce, Alex},
  booktitle={Proceedings of the IEEE/ACM 2nd International Workshop on Emerging Trends in Software Engineering for Blockchain (WETSEB)},
  pages={8--15},
  year={2019},
}

@misc{solhint,
  author = {Protofire},
  title = {Solhint: An Open Source Project for Linting Solidity Code},
  year = {2025},
  publisher = {GitHub},
  journal = {GitHub repository},
  howpublished = {GitHub repository, \url{https://github.com/protofire/solhint}},
  note = {version 0.6.0, Accessed: 2026-1-9}
}

@inproceedings{hu2023large,
  title={Large language model-powered smart contract vulnerability detection: New perspectives},
  author={Hu, Sihao and Huang, Tiansheng and {\.I}lhan, Fatih and Tekin, Selim Furkan and Liu, Ling},
  booktitle={Proceedings of the 5th IEEE International Conference on Trust, Privacy and Security in Intelligent Systems and Applications (TPS-ISA)},
  pages={297--306},
  year={2023},
}

@misc{aderyn,
  author = {Cyfrin},
  title = {Aderyn: A Rust-based Solidity Static Analyzer},
  year = {2026},
  publisher = {GitHub},
  howpublished = {GitHub repository, \url{https://github.com/Cyfrin/aderyn}},
  note = {Version 0.6.8, Accessed: 2026-01-17}
}

@inproceedings{luu2016making,
  title={Making smart contracts smarter},
  author={Luu, Loi and Chu, Duc-Hiep and Olickel, Hrishi and Saxena, Prateek and Hobor, Aquinas},
  booktitle={Proceedings of the 2016 ACM SIGSAC conference on computer and communications security (CCS)},
  pages={254--269},
  year={2016},
  note = {We utilized the Oyente+ fork (\url{https://github.com/smartbugs/oyente_plus}) for compatibility with modern Solidity.}
}

@article{hu2026effective,
  title={An Effective and Cost-Efficient Agentic Framework for Ethereum Smart Contract Auditing},
  author={Hu, Xiaohui and Chan, Wun Yu and Shi, Yuejie and Sun, Qumeng and Wang, Wei-Cheng and Wu, Chiachih and Wang, Haoyu and He, Ningyu},
  journal={arXiv preprint arXiv:2601.17833},
  year={2026}
}

@inproceedings{perkovic2024hallucinations,
  title={Hallucinations in llms: Understanding and addressing challenges},
  author={Perkovi{\'c}, Gabrijela and Drobnjak, Antun and Boti{\v{c}}ki, Ivica},
  booktitle={Proceedings of the 47th MIPRO ICT and electronics convention (MIPRO)},
  pages={2084--2088},
  year={2024},
  organization={IEEE}
}

@article{nelatoori2025toxic,
  title={Toxic comment classification and rationale extraction in code-mixed text leveraging co-attentive multi-task learning},
  author={Nelatoori, Kiran Babu and Kommanti, Hima Bindu},
  journal={Language Resources and Evaluation},
  volume={59},
  number={1},
  pages={161--190},
  year={2025},
  publisher={Springer}
}

@article{wang2025knowledge,
  title={Knowledge-enhanced recommendation via dynamic co-attention and high-order connectivity},
  author={Wang, Dan-Dong and Min, Fan},
  journal={International journal of machine learning and cybernetics},
  volume={16},
  number={2},
  pages={919--930},
  year={2025},
  publisher={Springer}
}

@misc{chen2025m3embeddingmultilingualitymultifunctionalitymultigranularity,
      title={M3-Embedding: Multi-Linguality, Multi-Functionality, Multi-Granularity Text Embeddings Through Self-Knowledge Distillation}, 
      author={Jianlv Chen and Shitao Xiao and Peitian Zhang and Kun Luo and Defu Lian and Zheng Liu},
      year={2025},
      eprint={2402.03216},
      archivePrefix={arXiv},
      primaryClass={cs.CL},
      url={https://arxiv.org/abs/2402.03216}, 
}

@inproceedings{xie2015holistically,
  title={Holistically-nested edge detection},
  author={Xie, Saining and Tu, Zhuowen},
  booktitle={Proceedings of the IEEE international conference on computer vision (ICCV)},
  pages={1395--1403},
  year={2015}
}

@inproceedings{drossos2025automated,
  title={Automated Market Makers: Toward More Profitable Liquidity Provisioning Strategies},
  author={Drossos, Thanos and Kirste, Daniel and Kannengie{\ss}er, Niclas and Sunyaev, Ali},
  booktitle={Proceedings of the 40th ACM/SIGAPP Symposium on Applied Computing (SAC)},
  pages={358--365},
  year={2025}
}

@inproceedings{sikder2025efficient,
  title={Efficient Adaptation of Large Language Models for Smart Contract Vulnerability Detection},
  author={Sikder, Fadul and Lei, Yu and Ji, Yuede},
  booktitle={Proceedings of the 21st International Conference on Predictive Models and Data Analytics in Software Engineering (PROMISE)},
  pages={65--74},
  year={2025}
}

@inproceedings{ma2025combining,
  title={Combining Fine-Tuning and LLM-Based Agents for Intuitive Smart Contract Auditing with Justifications},
  author={Ma, Wei and Wu, Daoyuan and Sun, Yuqiang and Wang, Tianwen and Liu, Shangqing and Zhang, Jian and Xue, Yue and Liu, Yang},
  booktitle={Proceedings of the 47th International Conference on Software Engineering (ICSE)},
  pages={1742--1754},
  year={2025},
  organization={IEEE}
}

@article{wu2025security,
  title={Security vulnerabilities in ethereum smart contracts: A systematic analysis},
  author={Wu, Jixuan and Xie, Lei and Li, Xiaoqi},
  journal={arXiv preprint arXiv:2504.05968},
  year={2025}
}

@article{yang2025multi,
  title={A Multi-Layered Security Analysis of Blockchain Systems: From Attack Vectors to Defense and System Hardening},
  author={Yang, Yuhuan and Ye, Shipeng and Li, Xiaoqi},
  journal={arXiv preprint arXiv:2504.09181},
  year={2025}
}

@article{bu2025smartbugbert,
  title={Smartbugbert: Bert-enhanced vulnerability detection for smart contract bytecode},
  author={Bu, Jiuyang and Li, Wenkai and Li, Zongwei and Zhang, Zeng and Li, Xiaoqi},
  journal={arXiv preprint arXiv:2504.05002},
  year={2025}
}

@article{wu2025exploring,
  title={Exploring vulnerabilities and concerns in solana smart contracts},
  author={Wu, Xiangfan and Xing, Ju and Li, Xiaoqi},
  journal={arXiv preprint arXiv:2504.07419},
  year={2025}
}

@article{ding2025comprehensive,
  title={A Comprehensive Study of Exploitable Patterns in Smart Contracts: From Vulnerability to Defense},
  author={Ding, Yuchen and Peng, Hongli and Li, Xiaoqi},
  journal={arXiv preprint arXiv:2504.21480},
  year={2025}
}

@article{li2025beyond,
  title={Beyond the Hype: A Large-Scale Empirical Analysis of On-Chain Transactions in NFT Scams},
  author={Li, Wenkai and Li, Zongwei and Li, Xiaoqi and Zhang, Chunyi and Zhang, Xiaoyan and Zhang, Yuqing},
  journal={arXiv preprint arXiv:2512.01577},
  year={2025}
}

@article{zhang2025penetration,
  title={Penetration testing for system security: Methods and practical approaches},
  author={Zhang, Wei and Xing, Ju and Li, Xiaoqi},
  journal={arXiv preprint arXiv:2505.19174},
  year={2025}
}

@article{zhu2024sybil,
  title={Sybil attacks detection and traceability mechanism based on beacon packets in connected automobile vehicles},
  author={Zhu, Yaling and Zeng, Jia and Weng, Fangchen and Han, Dan and Yang, Yiyu and Li, Xiaoqi and Zhang, Yuqing},
  journal={Sensors},
  volume={24},
  number={7},
  pages={2153},
  year={2024},
  publisher={MDPI}
}

\appendix

\end{document}